\documentclass[sn-basic]{sn-jnl}

\jyear{2021}%

\theoremstyle{thmstyleone}%
\theoremstyle{thmstyletwo}%

\theoremstyle{thmstylethree}%
\usepackage{amsfonts}

\begin{document}

\title[Reversible Jump Attack to Textual Classifiers with Modification Reduction]{Reversible Jump Attack to Textual Classifiers with Modification Reduction}


\author[1]{\fnm{Mingze} \sur{Ni}}\email{mingze.ni@student.uts.edu.au}

\author[2]{\fnm{Zhensu} \sur{Sun}}\email{sunzhs@shanghaitech.edu.cn}

\author*[1]{\fnm{Wei} \sur{Liu}}\email{wei.liu@uts.edu.au}

\affil[1]{\orgdiv{School of Computer Science}, \orgname{University of Technology Sydney}, \orgaddress{\street{15 Broadway}, \city{Sydney}, \postcode{2007}, \state{NSW}, \country{Australia}}}
\affil[2]{\orgdiv{School of Information Science and Technology}, \orgname{ShanghaiTech University}, \orgaddress{\street{393 Middle Huaxia Road}, \city{Shanghai}, \postcode{201210}, \state{Shanghai}, \country{China}}}

\newcommand{\su}[1]{{\color{pink}Zhensu: #1}}


\abstract{
Recent studies on adversarial examples expose  vulnerabilities of natural language processing (NLP) models. Existing techniques for generating adversarial examples are typically driven by deterministic hierarchical rules that are agnostic to the optimal adversarial examples, a strategy that often results in adversarial samples with a suboptimal balance between magnitudes of changes and attack successes. To this end, in this research we propose two algorithms, Reversible Jump Attack (RJA) and Metropolis-Hasting Modification Reduction (MMR), to generate highly effective adversarial examples and to improve the imperceptibility of the examples, respectively. RJA utilizes a novel randomization mechanism to enlarge the search space and efficiently adapts to a number of perturbed words for adversarial examples. With these generated adversarial examples, MMR applies the Metropolis-Hasting sampler to enhance the imperceptibility of adversarial examples. Extensive experiments demonstrate that RJA-MMR outperforms current state-of-the-art methods in attack performance, imperceptibility, fluency and grammar correctness.
}




\maketitle

\section{Introduction}
NLP models are known to be vulnerable in various applications, including machine translation \citep{ni2022attacking,cheng2020seq2sick,morphin2020tan}, sentiment analysis \citep{zang2020word, yang2021bigram}, and text summarization \citep{cheng2020seq2sick}. Attackers can exploit these weaknesses, creating adversarial examples that compromise the performance of targeted NLP systems. This growing susceptibility presents significant security challenges for AI models.

\par
Textual attacks on NLP models are classified into character \citep{iyyer2018sentence,ribeiro-etal-2018-sentence}, word \citep{alzantot2018generating,jia2019faga}, and sentence-level \citep{jia-liang} attacks. Character-level attacks are easily countered due to noticeable misspellings \citep{ebrahimi2017hotflip}, while sentence-level attacks often yield complex, hard-to-read text \citep{gan2019improving}. Word-level attacks are gaining preference for their effectiveness and subtlety, as they involve replacing words with carefully chosen substitutes \citep{Zhang2020AdversarialAO,garg2020bae,li2020bertattack}. Consequently, our focus is on conducting word-level adversarial attacks.
\par
\begin{figure}[t]
    \centering
    \includegraphics[width=\textwidth]{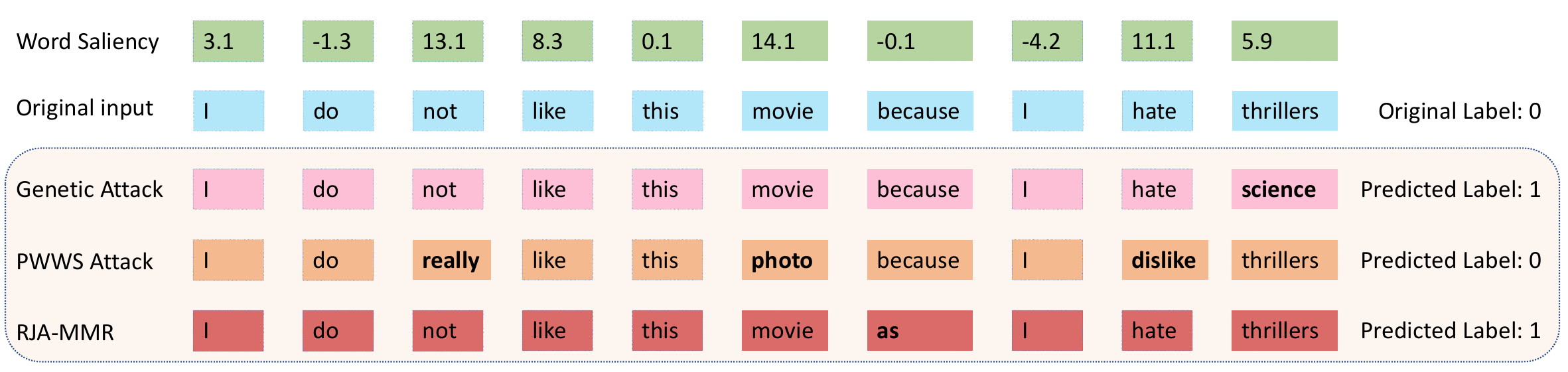}
    \caption{An illustrating example to show attack performances of optimizing attack (genetic attack), PWWS attack, and the proposed method RJA-MMR, where label ``0'' represents negative sentiment and ``1'' represents positive sentiment. The substitutions for different attack methods are bold. Genetic attack sacrifices too much semantics by changing ``thrillers" to ``science", while PWWS fails to fool the model and makes many ineffective modifications. The proposed method, RJA-MMR, makes a successful attack with only one word changed.}
    \label{fig: intro example}
\end{figure}

Crafting optimal adversarial examples involves navigating the interplay of successful attacks, controlled imperceptibility. The predominant strategies for this can be classified into \textbf{optimization algorithms} and \textbf{hierarchical search methods}. Within the realm of optimization, Genetic Attack (GA) \citep{alzantot2018generating,jia2019faga} and Particle Swarm Optimization (PSO) \citep{zang2020word} stand out as evolutionary approaches, focusing on optimizing attack effectiveness within embedding spaces and sememe-based thesauri, respectively. However, these methods face two primary challenges: 1) low efficiency in the optimization process due to the expansive search space, such as GloVe \citep{pennington2014glove}, and 2) Compromised semantic integrity, as even synonym-based word substitutions can cause sentence-level semantics inconsistency. On the other hand, Hierarchical search crafts adversarial examples by orderly substituting words based on word saliency rank (WSR) \citep{ren2019pwws,li2021clare,yang2021bigram}. It first identifies target words using WSR, then employs a Masked Language Model or thesaurus for substitutions. These hierarchical attacking methods have several drawbacks: 1) the first drawback of this approach is the difficulty of presetting the number of perturbed words (NPW) for large datasets with many tokens since the optimal NPW varies with different target texts \citep{tradeoff}; 2) the WSR-based methods will significantly reduce the searching domain by only attacking the combination of victim words ordered by the WSR. For a clear illustration, Fig \ref{fig: intro example} showcases the drawbacks of optimization-based GA and hierarchical PWWS attacks. GA’s replacement of `thriller' with `science' sacrifices semantic quality, while PWWS, despite altering three words, fails to fool the classifier.

\par
To address the above problems, we propose two novel black-box and word-level antagonistic algorithms: Reversible Jump Attacks (RJA) and MH Modification Reduction (MMR). For RJA, we employ the Reversible Jump sampler (RJS) and propose three variables from a target distribution: the number of perturbed words (NPW), victim words, and substitutions from Masked Language Models (MLM) and HowNet \citep{dong2003hownet}. The target distribution for RJS for evaluating the quality of the adversarial candidates is regularized by a strong penalty of semantic (dis)similarity. The NPW can be cross-dimensionally searched via RJS to adjust for different textual inputs according to their word saliency and overall performance. Given these three factors,  adversarial candidates are only accepted based on an acceptance probability from RJS. By running such a process iteratively, we will obtain the successful candidates with the highest semantic similarity. Therefore, RJA efficiently searches threat-level attacks inside a domain larger than WSR without presetting an NPW and sacrificing much semantics for imperceptibility.
\par
The other algorithm is Metropolis-Hasting Modification Reduction (MMR) which tends to restore the manipulations from RJA (i.e., reverse back to the original words) and then update the existing substitutions to maintain the attacking performance. Specifically, given an adversarial candidate, MMR first stochastically proposes a new candidate by restoring the attacked words. It applies a customized acceptance probability, calculated by comparing the overall performance between the new and current candidates, to determine the acceptance of the new candidate. After restoring some attacked words, MMR uses MH algorithm to update the substitutions of the current attacked words to preserve the attacking performance. By combining RJA and MMR, we proposed an integrated RJA-MMR as our final model. Specifically, RJA utilizes a Reverse Jump sampler \citep{Green1995ReversibleJM}, a Markov Chain Monte Carlo (MCMC) family member, to sample the dimensional jumping vectors to perform a cross-dimensional search for the optimal attacking performance constrained by semantic similarity. Intuitively, RJA and MMR agree on attacking performance improvement but disagree on NPW. By iteratively running these two antagonistic algorithms, attackers can boost the attack performance with only a small number of perturbations. The attack performance is illustrated by an example in Fig \ref{fig: intro example}, where RJA-MMR outperforms the optimizing attack (Genetic attack) and hierarchical attack (PWWS).
\par
Our main contributions from this work are as follows:
\begin{itemize}
    \item We design a highly effective adversarial attack method, Reversible Jump Attack (RJA), which utilizes the Reversible Jump algorithm to generate adversarial examples with an adaptive number of perturbed words. The algorithm enables our attack method to have an enlarged search domain by jumping across the dimensions.
    \item We propose Metropolis-Hasting Modification Reduction (MMR), which applies Metropolis-Hasting (MH) algorithm to construct an acceptance probability and use it to restore the attacked victim words to improve the imperceptibility with attacking performance reserved. MMR is functional with RJA and empirically proven effective in the adversarial examples generated by other attacking algorithms.
    \item We evaluate our attack method on real-world public datasets. Our results show that methods achieved the best performance in terms of attack performance, imperceptibility and examples' fluency.
\end{itemize}
\par

The rest of this paper is structured as follows. We first review adversarial attacks for NLP models and the Markov Chain Monte Carlo methods in NLP in Section \ref{related work}. Then we detail our proposed method in Section \ref{methodology}. We evaluate the performance of the proposed method through empirical analysis in Section \ref{experiments}. We conclude the paper with suggestions for future work in Section \ref{sec13}.

\section{Related Work}\label{related work}
This section reviews the literature on word-level textual attacks and MCMC sampling in NLP. 
\subsection{Word-level Attacks to Classifiers}
An increasing amount of effort is devoted to generating better textual adversarial examples with various attack models. Character-level attacks \citep{liang2018, ebrahimi2017hotflip} use misspellings to attach the victim classifiers; however, these attacks can often be defended by a spell checker. At the same time, sentence-level attacks \citep{iyyer-sentence-2018-adversarial,zou-sentence-2020-reinforced} pose threats to the classifier via inserting, removing, and paraphrasing sentences or pieces of sentences to the original input, while it's difficult for the generated text to maintain the imperceptibility \citep{li2021clare}. Word-level attacks pose non-trivial threats to NLP models by locating important words and manipulating them for targeted or untargeted purposes. Such attacks are broadly regarded as the optimal unit of attacks \citep{jia-liang}.

\subsubsection{Gradient-based Word-level Attacks}
With the help of an adopted fast gradient sign method (FGSM) \citep{Goodfellow2015fgsm}, Papernot et al.\citep{PapernotMSH16} were the first to generate word-level adversarial examples to classifiers. While their attack was able to fool the classifiers, their word-level manipulations significantly affected the original meaning. In \cite{liang2018}, the authors proposed to attack the target model by inserting Hot Training Phrases (HTPs) and modifying or removing the Hot Sample Phrases (HSPs), where HTPs and HSPs are calculated based on the gradient with respect to words from the input. Similar to Liang, \cite{samanta2018generating} utilizes the embedding gradient to determine the important words. Then hierarchical-driven rules together with hand-crafted word-level synonyms and character-level typos were designed. Notably, while the textual data is naturally discrete and more perceptible than image data, many gradient-based textual attacking methods inherited from computer vision are not effective enough, which leaves textual attack a challenging problem. 
\subsubsection{Non-gradient-based Word-level Attacks}
Alzantot et al.\citep{alzantot2018generating} transferred the domain of adversarial attacks to an optimization problem by formulating a customized objective function. With genetic optimization, they generate the adversarial examples by sampling the qualified genetic 'son' generations that break out the encirclement of the semantic threshold. However, the genetic algorithm can be low efficient. Since word embedding space is sparse, performing natural selection for languages in such a space can be computationally expensive. Jia \citep{jia2019faga} proposed a faster version of Alzantot's adversarial attacks by shrinking the search space, which accelerates the process of evolving in genetic optimization. Although Jia has greatly reduced the computational expense of genetic-based optimization algorithms, the optimizing processes inside word embedding space, such as GloVe \citep{pennington2014glove} and Word2Vec \citep{Mikolov2013word2vect}, are still not efficient enough. To ease the searching process, embedding-based algorithms have to use a counter-fitting method to post-process attacker’s vectors to accelerate the searching speed \citep{mrkvsic2016counter}. Compared with the word embedding method, utilizing well-organized linguistic thesaurus, e.g., synonym-based WordNet \citep{miller1990wordnet} and sememe-based HowNet \citep{dong2003hownet}, is a simple and easy implementation. Ren \citep{ren2019pwws} sought synonyms based on WordNet synsets and ranked word replacement order via probability-weighted word saliency (PWWS). Zang \citep{zang2020word} and Yang \citep{yang2021bigram} both
manifested that the sememe-based HowNet can
provide more substitute words via Particle Swarm Optimization (PSO) and an adaptive monotonic heuristic search to determine which group of words should be attacked. In addition, some recent studies utilized masked language models (MLM), such as BERT \citep{Devlin2019BERTPO} and RoBERTa \citep{Liu2019RoBERTaAR}, to generate contextual perturbations \citep{li2020bertattack,garg2020bae}. The pre-trained MLMs can ensure the predicted token correctly fits the sentence grammar but cannot preserve semantics. 
\subsection{Markov Chain Monte Carlo in NLP}
Markov chain Monte Carlo (MCMC) \citep{metropolis1953MH}, a statistically generic method for approximate sampling from an arbitrary distribution, can be applied in a variety of fields, such as optimization \citep{rubinstein1999crossoptimization}, machine learning \citep{fan2018rectangular}, quantum simulation \citep{haase2021quantum} and icing models \citep{herrmann1986ising}. The main idea is to generate a Markov chain whose equilibrium distribution is equal to the target distribution \citep{kroesehandbook}. There exist various algorithms for constructing chains, including the Gibbs sampler, Reversible Jump sampler \citep{green1995reversible}, and Metropolis-Hasting (MH) algorithm \citep{metropolis1953MH}. To get models capable of reading, deciphering, and making sense of human languages, NLP researchers apply MCMC to many downstream tasks, such as text generation and sentimental analysis. For text generation, Kumagai \citep{Kumagai2016HumanlikeNL} proposes a probabilistic text generation model which generates human-like text by inputting semantic syntax and some situational content. Since human-like text requests grammarly correct word alignment, they employed Monte Carlo Tree Search to optimize the structure of the generated text. In addition, Harrison \citep{harrison2017toward} presents the application of MCMC for generating a story, in which a summary of movies is produced by applying recurrent neural networks (RNNs) to summarize events and directing the MCMC search toward creating stories that satisfy genre expectations. For sentimental analysis, Kang \citep{Kang2011SamplingLE} applies the Gibbs sampler to the Bayesian network, a network of connected hidden neurons under prior beliefs, to extract the latent emotions. Specifically, they apply the Hidden Markov models to a hierarchical Bayesian network and embed the emotional variables as the latent variable of the Hidden Markov model. 

\subsubsection{Metropolis-Hasting and Reversible Jump Samplers}
The Metropolis-Hasting (MH) \citep{metropolis1953MH} algorithm is a classical Markov chain Monte Carlo sampling approach. Given the stationary distribution $f(\mathbf{z})$ and transition proposal $q(\mathbf{z}'|\mathbf{z})$, the MH algorithm can generate desirable examples from $f(\mathbf{z})$. Specifically, at each iteration, a new state $\mathbf{z'}$ will be proposed given the current state $\mathbf{z}$ based on a transition function $q(\mathbf{z}'|\mathbf{z})$. The MH algorithm is based on a ``trial-and-error" strategy by defining an acceptance probability $\alpha(\mathbf{z'}\vert \mathbf{z})$ as following:
\begin{align}
    \alpha(\mathbf{z'}\vert \mathbf{z})=\min \left\{\frac{f(\mathbf{z'}) q(\mathbf{z} \mid \mathbf{z'})}{f(\mathbf{z}) q(\mathbf{z'} \mid \mathbf{z})}, 1\right\} \label{eqt: mh original}
\end{align}
to decide whether the new state $\mathbf{z}'$ is accepted or rejected. 
\par

MCMC can also be applied to sample variational dimension sampling. Reversible Jump samplers (RJS) \citep{Green1995ReversibleJM} is a variation of MCMC algorithms specifically designed to sample from target distributions that contain vectors with different dimensions. Due to such a property, RJS can be applied to variable selection \citep{fan2011reversible}, dimension reduction \citep{rincent2017optimization}, and cross-dimensional optimization \citep{kroesehandbook}. Unlike the MH algorithm, RJS requests an additional transition item for proposing the new dimensions. The formulation of the acceptance probability of RJS is below:
\begin{align}
\displaystyle \alpha(\mathbf{z'}_{(m')}|\mathbf{z}_{(m)})=\min \left\{\frac{f(\mathbf{z'}_{(m')}) q(\mathbf{z}_{(m)} \mid \mathbf{z'}_{(m')})}{f(\mathbf{z}_{(m)}) q(\mathbf{z'}_{(m')} \mid \mathbf{z}_{(m)})}, 1\right\} \label{eqt: ori accept}
\\
    q\left(\mathbf{z'}_{(m')}|\mathbf{z}_{(m)}\right)=p\left(\mathbf{z'}_{(m')}|m', \mathbf{z}_{(m)}\right) p\left(m'|\mathbf{z}_{(m)}\right),\label{eqt: transition item}
\end{align} where $m$ denotes the dimensions of the vector $\mathbf{z}_{(m)}$, $q\left(\mathbf{z'}_{(m')}|\mathbf{z}_{(m)}\right)$ in Eq. \ref{eqt: transition item} illustrates the new transition function and $p\left(m'|\mathbf{z}_{(m)}\right)$ is the dimensional transition item. Comparing the acceptance probabilities of MH (Eq\ref{eqt: mh original}) and RJS (Eq\ref{eqt: ori accept}) reveals that RJS is more effective than MH in handling dimensional variations and sampling parameters of unknown dimensions. Since making adversarial would be a typical situation of dimension variation due to number of perturbed words (NPW), we believe that attacks based RJS is expected to achieve better performance than the literature based on MH \citep{MHA}.

\par
\subsubsection{Adversarial Attack via MCMC}
Despite the applications in NLP, the MCMC can be applied to adversarial attacks on NLP models. \cite{MHA} has successfully applied MH sampling to generate fluent adversarial examples for natural language by proposing gradient-guided word candidates. Specifically, they proposed both black-box and white-box attacks, and for black-box attacks, they perform removal, insertion and replacement by the words chosen from the pre-selector candidates set, but the empirical studies indicate these candidates are not efficient and effective for attacking. As for the white-box attacks, the gradient of the victim model is introduced to score the pre-selector candidates set, which successfully improves the attacking performance. However, the white-box setting is not practical in the real world, as attackers do not have access to the gradient and structure of the victim models. In addition, MHA successfully improved the language quality in terms of fluency, but the imperceptibility of the generated examples, especially in the modification rate, cannot be optimized.

\section{Imperceptible Adversarial Attack via Markov Chain Monte Carlo}\label{methodology}
In this section, we will detail our proposed method, RJA-MMR, the Reversible Jump attacks (RJA) with Metropolis-Hasting Modification Reduction (MMR).
\subsection{Problem Formulation and Notaition}
Given a pre-trained text classification model, which maps from feature space $\mathcal{X}$ to a set of classes $\mathcal{Y}$, an adversary aims to generate an adversarial document $\mathbf{x^*}$ from a legitimate document $x\in \mathcal{X}$ whose ground truth label is $y\in \mathcal{Y}$, so that $F(\mathbf{x^*})\neq y$. The adversary also requires  $Sem(x,\mathbf{x^*}) \leq \epsilon$ for a domain-specific semantic similarity function $Sem(\cdot): \mathcal{X}\times \mathcal{X}\rightarrow (0,1)$, where the bound $\epsilon \in \mathbb{R}$ helps to ensure imperceptibility. In other words, in the context of text classification tasks, we use $Sem(x,\mathbf{x^*})$ to capture the semantic similarity between $x$ and $\mathbf{x^*}$. More details of the notation are illustrated in Table \ref{tab: notation}.
\begin{table}[t]
    \centering
    \caption{List of notations used in this research.}
    \begin{tabular}{p{3.2cm}|p{7.5
cm}}
    \toprule
         Notation & Description \\ \toprule
         $\mathcal{X}$ & Text sample space. \\ \midrule
         $\mathcal{Y}$ & Class space.  \\\midrule
         $D$& A dataset to be attacked.\\\midrule
         $x=[w_1,w_2, \ldots,w_n]$ & An input text with n words and $w_i$ is the $i$th word in the sequence. \\\midrule
         $\mathbf{x}$ & An adversarial candidate generated by RJA. \\\midrule
         $m$, $\mathbf{v}$, $\mathbf{s}$ & Three factors in adversarial sample generation: the number of perturbed words, victim words, and their substitutions, respectively. \\\midrule
         $\mathbb{G}$ & The set of substitution candidates.\\ \midrule
         $\mathbf{x}^{r}$ & The adversarial candidate generated in the restoring step of MMR.  \\\midrule
         $\mathbf{x}^{u}$ & The adversarial candidate generated in the updating step of MMR. \\\midrule
         $\mathbf{x}^{*}$ & The final optima adversarial example.  \\\midrule
         $I(w_i)$ & The saliency of the word $w_i$. \\ \midrule
         $T$ & The total number of iterations for RJA-MMR. \\\midrule
         $F(\cdot):\mathcal{X}\rightarrow{\mathcal{Y}}$ & The victim classifier.\\\midrule
         $Sem(\cdot): \mathcal{X}^2\rightarrow (0,1)$ & The function measuring the semantic similarity.\\\midrule
         $p(\mathbf{x}_{t+1}|\mathbf{x}_{t}):\mathcal{X}\rightarrow(0,1)$ & The transition function from state $\mathbf{x}_{t}$ to $\mathbf{x}_{t+1}$.\\\midrule
         $\pi(x): \mathcal{X}\rightarrow (0,1)$& Target distribution.\\\midrule
         $\alpha(\mathbf{x}_{t+1}|\mathbf{x}_{t}): \mathcal{X}\rightarrow (0,1)$ & The acceptance probability.  \\
    \bottomrule
    \end{tabular}
    \label{tab: notation}
\end{table}

\begin{figure}[ht]
    \centering
    \includegraphics[width=\textwidth]{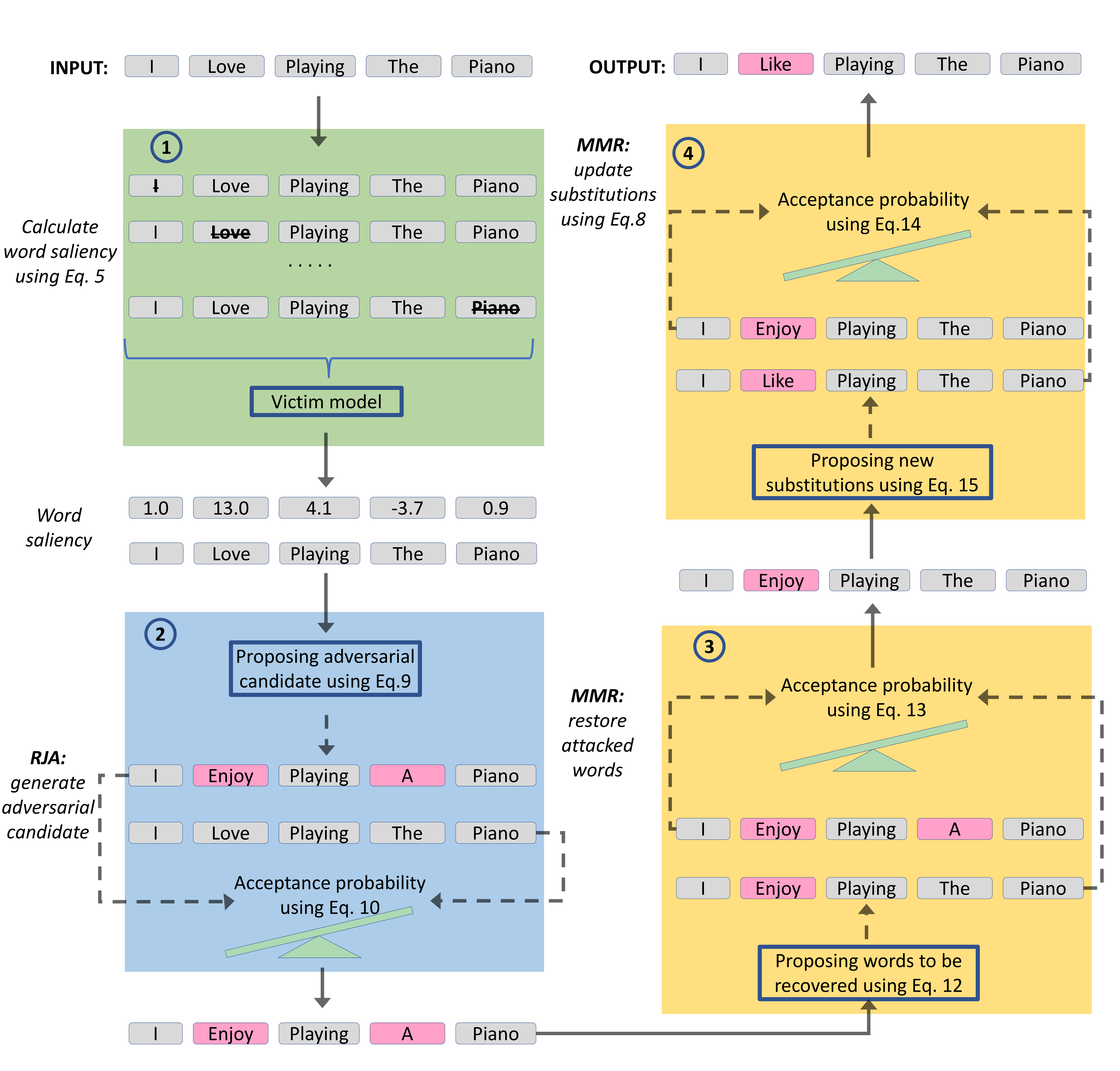}
    \caption{The workflow of our RJA-MMR. In this example, HAA generates an adversarial example with one word perturbed to attack a sentimental classifier with two labels (positive and negative). The block \textcircled{1} shows the calculation of word saliency. After obtaining the word saliency, we perform RJA in block \textcircled{2} which reflects the lines 4-15 in Algorithm \ref{algo: RJA}. After RJA, we perform the two steps, restoring and updating MMR in block \textcircled{3} and \textcircled{4}, respectively. The block \textcircled{3} and \textcircled{4} are illustrated in lines 4-10 and lines 11-18 in Algorithm \ref{algo: MMR}, respectively.}
    \label{fig:diagram}
\end{figure}

\subsection{Reversible Jump Attack}
This section details our proposed Reversible Jump Attack (RJA) which generates adversarial examples under semantic regularisation. Let $D=\{(x_1,y_1),(x_2,y_2),\ldots,(x_N,y_N)\}$ denote a dataset with $N$ data samples, where $x$ and $y$ are the input text and its corresponding class. Given the input text $x=[w_1,\ldots,w_i, \ldots, w_{n}]$ with $n$ words, we denote an adversarial candidate of RJA as $\mathbf{x}$ and denote the final chosen adversarial example as $\mathbf{x}^*$.
\par
RJA, unlike traditional methods, treats the number of perturbed words (NPW) as a variable in the sampling process, not a preset value. Utilizing the Reversible Jump Sampler, RJA conditionally samples NPW, victim words, and their substitutions. The approach involves a \textbf{transition function} that proposes adversarial candidates, evaluated against a \textbf{target distribution} focusing on attack effectiveness and semantic similarity (Eq. \ref{eqt: ori accept}). This process iteratively refines the adversarial examples, guided by an \textbf{acceptance probability} mechanism.
\par
This section first presents the transition function (Section \ref{transition function}) and then elaborates on the acceptance probability (Section \ref{accept probability RJA}), which builds upon the transition function.

\subsubsection{Transition Function}\label{transition function}

To propose the adversarial candidates, we construct our transition function to sequentially propose the three compulsory factors of crafting a new adversarial candidate $\mathbf{x}_{t+1}$ given the current one $\mathbf{x}_t$: the NPW $m$, the victim words $\mathbf{v}=[v_1, \ldots,v_m]$, and the corresponding substitutions $\mathbf{s}=[s_1, \ldots,s_m]$, where the dimension of $\mathbf{v}$ and $\mathbf{s}$ is $m$. Before we detail the process of proposing these factors, we first introduce the concept of the word saliency. In this context, word saliency refers to the impact of the word $w_i$ on the output of the classifier and 
the transition function, if this word is deleted from the sentence. 
The word with a high saliency has a high impact on the classifier. Thus, associating more importance to high-saliency words can help the transition function efficiently propose a high-quality adversarial candidate. To calculate the word saliency, we use the changes of victim classifiers’ logits before and after deleting word $w_i$ to represent the saliency $I( w_i)$:
\begin{align}
    I( w_i)=F_{logit}(x)-F_{logit}(x\backslash w_i),\label{eqt: saliency}
\end{align}
where $F_{logit}(\cdot)$ is the classifier returning the logit of the correct class, and $x\backslash w_i=[w_1,\ldots ,w_{i-1},w_{i+1},\ldots, w_{n}]$ is the text with $w_i$ removed. We calculate the word saliency $I(w_i)$ for all $w_i \in x$ to obtain word saliency $\mathbf{I}(x)$. Calculating the word saliency is illustrated in Block \textcircled{1} of Fig \ref{fig:diagram}.
\par

Among the iterations of searching for victim words, assume the RJA adversarial candidate at iteration $t$ is $\mathbf{x}_t=(m_t,\mathbf{v}_t,\mathbf{s}_t)$ and the new adversarial candidate to be crafted is $\mathbf{x}_{t+1}=(m_{t+1},\mathbf{v}_{t+1},\mathbf{s}_{t+1})$, we propose the first factor, the NPW value $m_{t+1}$, by either adding or subtracting 1, i.e., $m_{t+1}\in \{m_{t}+1, m_{t}-1\}$. This set $\{m_{t}+1, m_{t}-1\}$ does not need to include $m_{t}$ because if the proposed state is rejected, $m_{t+1}$ will be retained as $m_{t}$, which means $m_{t}$ still remains as a possible state. Thus the transition function for the new NPW value $m_{t+1}$ can be formulated as a probability mass function as below:
\begin{align}
    p&(m_{t+1}| \mathbf{x}_{t})=
    \begin{cases}
        \displaystyle \frac{\exp(l_1)}{\exp(l_1)+\exp(l_2)}& m_{t+1}=m_{t}-1,\\[0.8em]
      \displaystyle \frac{\exp(l_2)}{\exp(l_1)+\exp(l_2)} & m_{t+1}=m_{t}+1,\\[0.8em]
    \end{cases} \label{eqt: propose m}\\[0.5em]
    & \text{where} \quad l_1 = \sum_{w_i \in \mathbf{v}_{t}} I(w_i), \quad   l_2 = \sum_{w_i \notin \mathbf{v}_{t}} I(w_i). \nonumber
\end{align}

Such a transition function can propose the new state $m_{t+1}\in \{m_{t}-1,m_{t}+1\}$ by referring to the proportion of the exponential on victim word saliency $l_1$ and unattacked word saliency $l_2$ overall word saliency exponential. Intuitively, if the saliency values of all attacked words are high, the probability of proposing to reduce one attacked word, $m_{t+1}=m_{t}-1$, is high, and vice versa. Concretely, to sample $m_{t+1}$ from such a transition function, we firstly draw a random number, $\eta \sim Unif(0,1)$; and if $\eta$ is less than the probability of sampling $m_{t+1}=m_{t}-1$, i.e., $\eta<\frac{\exp(l_1)}{\exp(l_1)+\exp(l_2)}$, then $m_{t+1}=m_{t}-1$, otherwise $m_{t+1}=m_{t}+1$. Unlike hierarchical attacks, which deterministically perturb the words in the descending order of the word saliency, randomization is applied because of its two merits: 1) it overcomes the imprecision problem with the WSR (word saliency rank) mentioned in the preceding introduction section, and 2) it enlarges the search domain by proposing more combinations of attacked words than those in hierarchical searching. 

\par
After determining the number of perturbed words, we sample one target victim word $v_{tgt}$ (where ``tgt'' refers to ``target'') to be manipulated according to the newly sampled $m_{t+1}$. 
Specifically, for $m_{t+1}=m_{t}+1$, the target word $v_{tgt}$ is uniformly sampled from unattacked word set $x \backslash \mathbf{v}_{t}$, while for $m_{t+1}=m_{t}-1$ the target word $v_{tgt}$ is uniformly drawn from attacked words set $\mathbf{v}_{t}$ then the selected words will be restored to the original words. The transition function of sampling the target victim word $v_{tgt}$ is thus formulated as:
\begin{align}
\displaystyle
    p(v_{tgt}|\mathbf{x}_{t},m_{t+1})&=
    \begin{cases}
        \frac{1}{m_{t}}\quad v_{tgt}\in \mathbf{v}_{t}& \text{       if } m_{t+1}=m_{t}-1,\\[0.8em]
        \frac{1}{n-m_{t}}\quad v_{tgt}\in \mathbf{x}\backslash\mathbf{v}_{t}& \text{   if }m_{t+1}=m_{t}+1.\\
    \end{cases} \label{eqt: propose v}
\end{align} 
\par
After the target word $v_{tgt} \in \mathbf{x}_{t}$ is selected, we search for a parsing-fluent and semantic-preserving substitution for $w_{tgt}$. Therefore, we uniformly draw a substitution $s_{tgt}$ for $v_{tgt}$ from the candidates set, which is the intersection (consensus) of candidates provided by Mask Language Models (MLMs) and Synonyms. Specifically, let $\mathcal{M}$ denote the MLM, and we mask the $v_{tgt}$ in $\mathbf{x}$ to construct a masked $\mathbf{x}_{mask}$ and feed the masked text into $\mathcal{M}$ to search for the parsing-fluent candidates. Instead of using the argmax prediction, we take the most possible $K$ words, which are the top $K$ words suggested by the logits from $\mathcal{M}$, to construct MLM candidates set $\mathbb{G}_{\mathcal{M}}=\{w_{\mathcal{M}}^{1}, \ldots,w_{\mathcal{M}}^{K}\}$. To keep semantically similar, we form a synonym set $\mathbb{G}_{syn}=\{w_{syn}^1, \ldots, w_{syn}^K\}$ from HowNet \citep{dong2010hownet} based thesauri such as OpenHowNet \citep{qi2019openhownet} and BabelNet \citep{BabelNet} These thesauri are context-aware and at the same time can provide more synonyms than common thesaurus such as WordNet \citep{Miller1992WordNetAL}. Since our objective is that the generated adversarial examples should be parsing-fluent and semantic-preserving, the substitution $s_{tgt}$ will be uniformly sampled from the intersection $\mathbb{G}=\mathbb{G}_{\mathcal{M}}\cap \mathbb{G}_{syn}$, which is illustrated in Eq. \ref{eqt: propose s}.
\begin{align}
    p(s_{tgt}|w_{tgt},m_{t+1},\mathbf{x}_{t})=\frac{1}{[\mathbb{G}]} \label{eqt: propose s}
\end{align}
where $\mathbb{G}=\mathbb{G}_{\mathcal{M}}\cap \mathbb{G}_{syn}$ and $[\mathbb{G}]$ is the cardinality of the set $\mathbb{G}$.
\par
By applying the Bayes rule to the Eq. \ref{eqt: propose m}, \ref{eqt: propose v} and \ref{eqt: propose s}, the final transition function is:
\begin{align}
p_{_{RJA}}\left(\mathbf{x}_{t+1}|\mathbf{x}_{t}\right)=p\left(m_{t+1}|\mathbf{x}_{t}\right)p\left(w_{tgt}|m_{t+1},\mathbf{x}_{t}\right)p\left(s_{tgt}|w_{tgt},m_{t+1},\mathbf{x}_{t}\right)\label{eqt: transition fucntion}
\end{align}
\par

\subsubsection{Acceptance Probability for RJA} \label{accept probability RJA}
Before we calculating the acceptance probability, we need to construct the \textbf{target distribution} for evaluating the performance. Specifically, we argue that a good adversarial example should achieve successful attacks while being kept semantically similar to the input text $x$. Therefore, we formulate the following equation as our target distribution:
\begin{align}
    \pi(\mathbf{x})=\frac{\left(1-F_{p}(\mathbf{x})\right)Sem\left(x,\mathbf{x}\right)}{C},\label{eqt: target rja}
\end{align}
where $Sem(x,\mathbf{x})$ represents the semantic similarity, which generally is implemented with the cosine similarity between sentence encodings from a pre-trained sentence encoder, such as USE \citep{cer2018universal}. $C=\sum_{\mathbf{x}\in \mathcal{X}}\left(1-F_{p}(\mathbf{x})\right)Sem\left(x,\mathbf{x}\right)$ is a positive normalizing factor to make $\sum_{\mathbf{x}\in \mathcal{X}}\pi(\mathbf{x})=1$ and $F_{p}(\cdot): \mathcal{X}\rightarrow (0,1)$ denotes the confidence of making right predictions where $\mathcal{X}$ represents text space. From Eq. \ref{eqt: target rja}, we can easily observe that the value from target distribution $\pi(\mathbf{x})$ will increase with the increase of the attacking performance measured by the confidence of making a wrong prediction $1-F_{p}(\mathbf{x})$, and semantic similarity $Sem(x,\mathbf{x})$.
\par
Given the \textbf{target distribution} in Eq. \ref{eqt: target rja} and \textbf{transition function} in Eq. \ref{eqt: transition fucntion}, we formulate the acceptance probability for RJA, $\alpha_{_{RJA}}(\mathbf{x}_{t+1}|\mathbf{x}_{t})$, as follows:
\begin{align}
    \displaystyle \alpha_{_{RJA}}(\mathbf{x}_{t+1}|\mathbf{x}_{t})=\min \left\{\frac{\pi(\mathbf{x}_{t+1}) p_{_{RJA}}(\mathbf{x}_{t}|\mathbf{x}_{t+1})}{\pi(\mathbf{x}_{t}) p_{_{RJA}}(\mathbf{x}_{t+1}|\mathbf{x}_{t})}, 1\right\}
    \label{eqt: rja accept}
\end{align}
After calculating $\alpha(\mathbf{x}_{t+1}|\mathbf{x}_{t})$, we sample a random number $\epsilon$ from a uniform distribution, $\epsilon \sim Uniform(0,1)$, if $\epsilon<\alpha(\mathbf{x}_{t+1}|\mathbf{x}_{t})$ we will accept $\mathbf{x}_{t+1}$ as the new state, otherwise the state will remain as $\mathbf{x}_{t}$. By running $T$ iterations, we obtain a set of adversarial candidates $\{\mathbf{x}_1,\mathbf{x}_2,\ldots \mathbf{x}_T\}$. We then choose the candidate which not only successfully fools the classifier but also preserves the most semantics as the final adversarial candidate $\mathbf{x}$. The process of RJA is illustrated in Algorithm \ref{algo: RJA} and block \textcircled{2} in Fig \ref{fig:diagram}.

\begin{algorithm}[t]
\label{algo: RJA}
\caption{Reversible Jump Attack (RJA)}
\DontPrintSemicolon
\KwInput{Input text: $x$, Number of iterations: $T$}
\KwOutput{Adversarial candidate $\mathbf{x}$}
$Adv\_set=[\quad]$\;
$\mathbf{x}_{0}=x$\;
\For{ t+1 in range($T$)}{
Sample $m_{t+1}$ given $\mathbf{x}_{t}$ with Eq. \ref{eqt: propose m} \;
Sample $\mathbf{s}_{t+1}$ given $\mathbf{x}_{t}$ and $m_{t+1}$ with Eq. \ref{eqt: propose v} \;
Sample $\mathbf{v}_{t+1}$ given $\mathbf{v}_{t}$, $m_{t+1}$ and $\mathbf{s+1}$ with Eq. \ref{eqt: propose s} \;
Craft  $\mathbf{x}_{t+1}$ with $m_{t+1}$, $\mathbf{s}_{t+1}$, $\mathbf{v}_{t+1}$ and $\mathbf{x}_{t-1}$,\;
Calculate the acceptance probability, $\alpha(\mathbf{x}_{t}|\mathbf{x}_{t-1})$ with Eq. \ref{eqt: rja accept}\;
Sample $\epsilon$ from Uniform distribution, Uniform(0,1)\;
\uIf{$\epsilon<\alpha(\mathbf{x}_{t}|\mathbf{x}_{t-1})$}
{
$\mathbf{x}_{t+1}=\mathbf{x}_{t+1}$\;
$Adv\_set=[Adv\_set,~\mathbf{x}_{t+1}]$\;
}
\Else{
$\mathbf{x}_{t+1}=\mathbf{x}_{t}$\;
$Adv\_set=[Adv\_set,~\mathbf{x}_{t+1}]$\;
}
\Return $Adv\_set$
}
Choose the candidate which successfully fools the classifier with lease semantic sacrifice as an adversarial example $\mathbf{x}$.\;
\Return Adversarial candidate $\mathbf{x}$

\end{algorithm}

\subsection{Modification Reduction with Metropolis-Hasting Algorithm}
Besides the success of tampering with the classifier and semantic preservation, the modification rate is also an important factor in evaluating the imperceptibility of adversarial examples. Generally, methods in the literature can generate effective adversarial examples; however, it was hard to guarantee the modification rate is optimally the lowest. To address this, we introduce the Metropolis-Hasting Modification Reduction (MMR), leveraging the Metropolis-Hasting (MH) algorithm to optimize the modification rate by exploring efficient yet minimal substitution combinations for a given adversarial candidate. MMR involves two steps, each employing the MH algorithm: 1) stochastically \textbf{restoring} some attacked words to create a less modified candidate and 2) \textbf{updating} all substitutions without altering the NPW, \(m\). These steps are detailed in Sections \ref{MMR restore} and \ref{MMR update} respectively.

\subsubsection{Restoring Attacked Words with MMR}\label{MMR restore}
The first step of MMR is probabilistically restoring some attacked words with MH algorithm to test the necessity of the current substitutions. Given an adversarial candidate $\mathbf{x}_t=(m_t, \mathbf{v}_t,\mathbf{s}_t)$ from iteration $t$ in RJA, we aim to generate an adversarial candidate $\mathbf{x}_t^{r}$ which is constructed by restoring some attacked words in $\mathbf{x}_t$. To sample the restored substitutions, we propose the probability mass function of selecting substitutions $s^r \in \{s_i,w_i\}$ in iteration $t$ as follows:
\begin{align}
    p(s^r|\mathbf{x}_t)&=
    \begin{cases}
        \displaystyle \frac{\exp(I(w_i))}{1+\exp(I(w_i))}& \text{       if } s^r = s_{i}\; (\text{continue to attack}),\\[0.8em]
        \displaystyle \frac{1}{1+\exp(I(w_i))} & \text{       if } s^r = w_{i} \;(\text{attack cancelled}),\\[0.8em]
    \end{cases} \label{eqt: propose m and s}\\[0.5em]
    p_{restore}(&\mathbf{x}^{r}_{t}|\mathbf{x}_t)=\prod_{s^r \in \mathbf{s}_t}p(s^r|\mathbf{x}_{t})
\end{align}
where $s^r=s_i$ denotes to continue the attack and $s^r=w_i$ denotes restoring the substitution to the original word $w_i$, respectively. The $\mathbf{x}^{r}_{t} $ is the proposed adversarial candidate with selected substitutions restored from $\mathbf{x}$. With such a probability mass function, the $s^r$ can be sampled by the same strategy of sampling as in Eq. \ref{eqt: propose m}. To further investigate the quality of such a candidate, we apply the target distribution, $\pi(\mathbf{x})$, in Eq. \ref{eqt: target rja} to construct the following acceptance probability:
\begin{align}
    \displaystyle \alpha_{restore}(\mathbf{x}^{r}_{t}|\mathbf{x}_t)=\min \left(\frac{\pi(\mathbf{x}^{r}_{t}) p_{restore}(\mathbf{x}_t|\mathbf{x}^{r}_{t})}{\pi(\mathbf{x}_t) p_{restore}(\mathbf{x}^{r}_{t}|\mathbf{x}_t)}, 1\right)
    \label{eqt: accept restore}
\end{align}
to decide whether the proposed adversarial candidate $\mathbf{x}^{r}_t$ should be accepted as the true candidate.

\subsubsection{Updating the Combination of Substitutions with MMR} \label{MMR update}
Having restored selected substitutions to obtain the adversarial candidate \( \mathbf{x}^{r}_{t} \) at the \( t \)-th iteration, we proceed to the second step: MMR updating. This step is designed to refine attack performance by altering substitution combinations without affecting the NPW, \( m_t \). We apply a methodology similar to the one in Eq. \ref{eqt: propose s} for sampling substitution combinations. In essence, the MMR updating utilizes the candidate proposing function (Eq. \ref{eqt: propose s}) to explore alternative substitutions for each attacked word, aiming for enhanced attack efficacy. The formulation for this update, leading to the next adversarial candidate \( \mathbf{x}^{u}_{t} \), is governed by the subsequent acceptance probability:
\begin{align}
    \alpha_{update}(\mathbf{x}^{u}_{t}|\mathbf{x}^{r}_{t})&=\min \left(\frac{\pi(\mathbf{x}^{u}_{t}) p_{update}(\mathbf{x}^{r}_{t}|\mathbf{x}^{u}_{t})}{\pi(\mathbf{x}^{r}_{t}) p_{update}(\mathbf{x}^{u}_{t}|\mathbf{x}^{r}_{t})}, 1\right) \label{eqt: accept update},\\[0.5em]
    p_{update}(\mathbf{x}^{u}_{t}|\mathbf{x}^{r}_{t})&=\prod_{s_i \in \mathbf{s}^{r}_t}p(s_i|w_{i}, m^{r}_{t},\mathbf{x}^{r}_{t}) \label{eqt: propose ss},
\end{align}
where $p(s_i|w_{i},m^{r}_{t},\mathbf{x}^{r}_{t})$ is identical to that in Eq. \ref{eqt: propose s}. 
\par
By iteratively running $T$ times MH algorithms for substitution restoring and updating with acceptance probabilities in Eq. \ref{eqt: accept restore} and Eq. \ref{eqt: accept update}, respectively, we can construct the adversarial set $\mathbb{X}'=\{\mathbf{x}^u_{t}\}^{T}_{t=1}$ and select the candidate with the highest semantic similarity among the successful candidates that fools the classifier as the final adversarial example $\mathbf{x}^{*}$. This proposed MMR algorithm will not only be applied to our RJA algorithm but also can help other attack methodologies reduce their modifications. The whole process of MMR is illustrated in Algorithm \ref{algo: MMR} and block \textcircled{3}-\textcircled{4} in Fig \ref{fig:diagram}.
\begin{algorithm}[t]
\caption{Metropolis-Hasting Modification Reduction (MMR)}
\KwInput{Adversarial candidate $\mathbf{x}=(m,\mathbf{v},\mathbf{s})$}
\KwOutput{The final adversarial example $\mathbf{x}^*$}
\label{algo: MMR}
\DontPrintSemicolon
$Adv\_set=[\quad]$\;
\For{ t in range($T$)}{
Fetch $\mathbf{x}_t$ from RJA in iteration $t$\;
Sample $\mathbf{x}^r_{t}$ to reduce the modifications with Eq. \ref{eqt: propose m and s} \;
Calculate the acceptance probability, $\alpha(\mathbf{x}_{t}|\mathbf{x})$ with Eq. \ref{eqt: accept restore}\;
Sample $\epsilon$ from Uniform distribution, Uniform(0,1)\;
\uIf{$\epsilon<\alpha(\mathbf{x}^r_{t}|\mathbf{x})$}
{
$\mathbf{x}^r_t=\mathbf{x}^r_t$\;
}
\Else{
$\mathbf{x}^r_t=\mathbf{x}$\;
}
Sample $\mathbf{x}^u_{t}$ to update the substitutions in $\mathbf{x}^r_t$ with Eq. \ref{eqt: propose ss} \;
Calculate the acceptance probability, $\alpha(\mathbf{x}^u_{t}|\mathbf{x}^r_{t})$ with Eq. \ref{eqt: accept update}\;
\uIf{$u<\alpha(\mathbf{x}^u_{t}|\mathbf{x}^r_{t})$}
{
$\mathbf{x}^u_t=\mathbf{x}^u_t$\;
Take $\mathbf{x}^u_t$ as RJA's input for next iteration\;
}
\Else{
$\mathbf{x}^u_t=\mathbf{x}^r_{t}$\;
Take $\mathbf{x}^u_t$ as RJA's input for next iteration\;
}
$Adv\_set=[Adv\_set,~\mathbf{x}^u_{t}]$\;
\Return $Adv\_set$
}
Choose the candidate with the least modification from $Adv\_set$ as the final adversarial example $\mathbf{x}^*$.\;
\Return The final adversarial example $\mathbf{x}^*$
\end{algorithm}

\section{Experiments and Analysis}\label{experiments}
In this section, we comprehensively evaluation the performance of our method against the current state of the art. Besides the main results (Sec. \ref{main results}) of attacking performance and imperceptibility, we also conduct experiments on ablation studies (Sec. \ref{ablation}), efficiency analysis (Sec. \ref{efficiency}), transferability (Sec. \ref{transferability}), target attacks (Sec. \ref{target}), performance front of defense mechanism (Sec. \ref{defense}), adversarial retraining (Sec. \ref{retraining}), part-of-speech (POS) preference (Sec. \ref{preference}) and scales of models for robustness(Sec. \ref{scale})

We evaluate the effectiveness our methods on three widely-used and publicly available benchmark datasets: AG’s News \citep{ag_news}, Emotion \citep{emotion}, SST2 \citep{sst2} and IMDB\citep{IMDB}. Specifically, AG’s News is a news classification dataset with 127,600 samples belonging to 4 topic classes, \textit{World, Sports, Business, Sci/Tech}. Emotion \citep{emotion} is a dataset with 20,000 samples and 6 classes, \textit{sadness, joy, love, anger, fear, surprise}. SST2 \citep{sst2} is a binary class (\textit{positive and negative}) topic dataset with 9,613 samples. The IMDB dataset \citep{IMDB}, comprising movie reviews from the Internet Movie Database, is predominantly utilized for binary sentiment classification, categorizing reviews into `positive' or `negative' sentiments. The details of these datasets can be found in Table \ref{tab: datasets and models}.

To ensure reproducibility, we provide the code and data used in our experiments in a GitHub repository\footnote{ \url{https://github.com/MingzeLucasNi/RJA-MMR}}.
 
\begin{table}[t]
\small
\caption{Datasets and accuracy of victim models before attacks.}
\centering
\begin{tabular}{ccccccc}
\toprule
Dataset &Size &Avg.Length & Class & Task      &  Model         &  Accuracy  \\
\midrule
\multirow{2}[0]{*}{AG's News}    &\multirow{2}[0]{*}{12,700}
&\multirow{2}[0]{*}{37.84}&\multirow{2}[0]{*}{4}    &\multirow{2}[0]{*}{\shortstack{News\\ topics}}&
BERT-C &  94\%    \\
~&~&~&~&~     &  TextCNN   &  90\%  \\
\midrule
\multirow{2}[0]{*}{Emotion}  &
\multirow{2}[0]{*}{20,000}  &
\multirow{2}[0]{*}{19.14}&\multirow{2}[0]{*}{6}    &\multirow{2}[0]{*}{\shortstack{Sentiment\\analysis}}&  BERT-C      &   97\%
\\
~  &~&~ &~&~&  TextCNN &   93\%  \\
\midrule
\multirow{2}[0]{*}{SST2} & 
\multirow{2}[0]{*}{9,613}  &
\multirow{2}[0]{*}{19.31}&\multirow{2}[0]{*}{2}    &\multirow{2}[0]{*}{\shortstack{Sentiment\\analysis}}
&BERT-C      &   91\%   \\
~&~&~&~&~ &  TextCNN   &  83\% \\

\midrule
\multirow{2}[0]{*}{IMDB} & 
\multirow{2}[0]{*}{50,000}  &
\multirow{2}[0]{*}{279.48}&\multirow{2}[0]{*}{2}    &\multirow{2}[0]{*}{\shortstack{Movie\\review}}
&BERT-C      &   93\%   \\
~&~&~&~&~ &  TextCNN   &  88\% \\

\bottomrule
\end{tabular}
\label{tab: datasets and models}
\end{table}

\subsection{Victim Models}
We apply our attack algorithm to two types of popular and well-performed victim models. The details of the models can be found below.
\subsubsection*{BERT-based Classifiers} To do convincing experiments, we choose three well-performed and popular BERT-based models, which we call BERT-C models (where the letter ``C'' represents ``classifier''),  pre-trained by Huggingface\footnote{\url{https://huggingface.co/}}. Due to the different sizes of the datasets, the structures of BERT-based classifiers are adjusted accordingly. The BERT classifier for AG's News is structured by the \textit{Distil-RoBERTa-base} \citep{Sanh2019DistilBERTAD} connected with two fully connected layers, and it is trained for 10 epochs with a learning rate of 0.0001. For the Emotion dataset, its BERT-C adopts another version of BERT, \textit{Distil-BERT-base-uncased} \citep{Sanh2019DistilBERTAD}, and the training hyper-parameters remain the same as BERT-C for AG's News. Since the SST2 dataset is relatively small compared with the other two models, the corresponding BERT classifier utilizes a small-size version of BERT, \textit{BERT-base-uncased} \citep{Devlin2019BERTPO}. As for the IMDB, we employ the \textit{Distil-BERT-base-uncased} for classification tasks. The test accuracy of these BERT-based classifiers before they are under attacks are listed in Table \ref{tab: datasets and models} and these models are publicly accessible\footnote{\url{https://huggingface.co/mrm8488/distilroberta-finetuned-age_news-classification}} \footnote{\url{https://huggingface.co/bhadresh-savani/distilbert-base-uncased-emotion}} \footnote{\url{https://huggingface.co/echarlaix/bert-base-uncased-sst2-acc91.1-d37-hybrid}} \footnote{\url{https://huggingface.co/lvwerra/distilbert-imdb}}.
\par
\subsubsection*{TextCNN-based models}The other type of victim model is TextCNN \citep{Kim2014ConvolutionalNN}, structured with a 100-dimension embedding layer followed by a 128-units long short-term memory layer. This classifier is trained 10 epochs by ADAM optimizer with parameters: learning rate $lr=0.005$, the two coefficients used for computing running averages of gradient and its square are set to be 0.9 and 0.999 $(\beta_1=0.9$, $\beta_2=0.999)$,  the denominator to improve numerical stability $\sigma=10^{-5}$. The accuracy of these TextCNN-base models is also shown in Table \ref{tab: datasets and models}.

\subsection{Baselines}
To evaluate the attacking performance, we use the TextAttack \citep{morris2020textattack} framework to deploy the following baselines:
\begin{itemize}
    \item AGA \citep{alzantot2018generating}: it uses the combination of restrictions on word embedding distance and language model prediction scores to reduce search space. As for the searching algorithm, it adopts a genetic algorithm, a popular population-based evolutionary algorithm.
    \item Faster Alzantot Genetic Algorithm (FAGA) \citep{jia2019faga}: it accelerates AGA by bounding the searching domain of genetic optimization.
    \item BERT-Base Adversarial Examples (BAE) \citep{garg2020bae}: it replaces and inserts tokens in the original text by masking a portion of the text and leveraging the BERT-MLM. 
    \item Metropolis-Hasting Attack (MHA) \citep{MHA}: it performs Metropolis-Hasting sampling, which is designed with the guidance of gradients, to sample the  examples from a pre-selector that generates candidates by using MLM.
    \item BERT-Attack (BA)\citep{li2020bertattack}: it takes advantage of BERT-MLM to generate candidates and attacked words by the static WSR descending order.
    \item Probability Weighted Word Saliency (PWWS) \citep{ren2019pwws}: it chooses candidate words from WordNet \citep{miller1990wordnet} and sorts word attack order by multiplying the word saliency and probability variation.    
    \item TextFooler (TF) \citep{Jin2020IsBR}: it ranks the important words with similar strategy with Eq. \ref{eqt: saliency}. With the important rank, the attacker prioritizes replacing them with the most semantically similar and grammatically correct words until the prediction is altered.
    \item Particle Swarm Optimization (PSO) \citep{zang2020word}: it selects word candidates from HowNet and employs the POS to find adversarial text. This method treats every sample as a particle whose location in the search space needs to be optimized.    
\end{itemize}

\subsection{Experimental Settings and Evaluation Metrics}
For our RJA and RJA-MMR, we use the Universal Sentence Encoder (USE) \citep{cer2018universal} to measure the sentence semantic similarity for target distribution in Eq. \ref{eqt: target rja}. We experiment to find $k=30$ substitution candidates and to find these candidates' substitutions, we use \textit{RoBERTa-large} \citep{Liu2019RoBERTaAR} as the MLM with WordPiece \citep{wordpiece} tokenizer for contextual infilling and utilize OpenHowNet \citep{qi2019openhownet} with NLTK \citep{nltk} tokenizer as the synonym thesaurus. For the sampling-based algorithms, MHA and the proposed methods (RJA, RJA-MMA), we set the maximum number of iterations $T$ to $1000$. 

We argue that the quality of adversarial examples is appraised with regard to three key facets: attacking performance, imperceptibility, and fluency. To measure these facets, we use the following five metrics to measure the performance of adversarial attacks:
\begin{itemize}
    \item Successful attack rate (SAR) is defined as the percentage of attacks where the adversarial examples make the victim models predict a wrong label.
    \item Modification Rate(Mod) is the percentage of modified tokens. Each replacement, insertion or removal action accounts for one modified token.
    \item Grammar Error (GErr) is measured by the absolute rate of increased grammatic errors in the successful adversarial examples, compared to the original text, where we use LanguageTool \citep{naber2003rule} to obtain the number of grammatical errors.
    \item Perplexity (PPL) denotes a metric used to evaluate the fluency of adversarial examples \citep{kann2018sentence,zang2020word}. The perplexity is calculated using small-sized GPT-2 with a 50k-sized vocabulary \citep{radfordlanguage}.
    \item Textual similarity (Sim) is measured by the cosine similarity between the sentence embeddings of the input and that of the adversarial sample. We encoded the two sentences with the universal sentence encoder (USE) \citep{cer2018universal}.
\end{itemize} SAR evaluates attack performance, while Mod and Sim measure imperceptibility. GErr and PPL assess language fluency.

\begin{landscape}
\begin{table}[ht]
\caption{Results on SAR, Mod, and Sim metrics among the baselines and proposed methods on different datasets. The best performance is in bold.}
\small
\centering
\begin{tabular}{llcccccccccccc}
\toprule
\multirow{2}{*}{Task} & \multirow{2}{*}{Method} & \multicolumn{3}{c}{AG News} & \multicolumn{3}{c}{Emotion} & \multicolumn{3}{c}{SST2}& \multicolumn{3}{c}{IMDB} \\ \cmidrule(lr){3-5} \cmidrule(lr){6-8} \cmidrule(lr){9-11} \cmidrule(lr){12-14}
                      &                         & SAR$\uparrow$ & Mod$\downarrow$ & Sim$\uparrow$ & SAR$\uparrow$ & Mod$\downarrow$ & Sim$\uparrow$ & SAR$\uparrow$ & Mod$\downarrow$ & Sim$\uparrow$ & SAR$\uparrow$ & Mod$\downarrow$ & Sim$\uparrow$ \\
\hline
\multirow{10}{*}{BERT-C} 
                      & BAE                    & 41.0 & 11.3 & 72   & 68.7 & 7.7   & 88  & 55.1 & 11.3 & 75 &66&6.9&\textit{89}\\
                      & AGA                    & 15.3 & 10.9 & 71   & 48.1 & 8.1   & 89  & 41.7 & 12.2 & 70 &71&6.6&\underline{90}\\
                      & FAGA                   & 27.9 & 14.6 & 75   & 78.2 & 9.8   & 89  & 77.1 & 17.2 & 69 &81&7.3&\textit{89}\\
                      & MHA                    & 41.2 & 14.2 & 62   & 77.1 & 10.2  & 87  & 84.5 & 17.2 & 72 &79.9&6.9&81\\
                      & BA                     & 47.7 & 17.4 & 73   & 85.9 & 9.2   & 82  & 77.2 & 13.5 & 74 &78.1&\underline{5.6}&87\\
                      & PWWS                   & 76.8 & 17.9 & 73   & 91.8 & 10.2  & 83  & 93.9 & 17.2 & 84 &\textit{99.0}&7.8&80\\
                      & TF                     & 89.3 & 21.6 & 77   & 90.1 & 10.2  & 82  & 90.4 & 15.3 & 80 &96.3&\underline{5.6}&81\\
                      & PSO                    & 93.4 & 21.6 & 69   & 94.3 & 12.0  & 87  & 96.6 & 17.2 & 81 &\underline{99.1}&6.3&80\\
                      & RJA                    & 95.1 & 11.1 & 77   & 97.1 & 8.5   & 88  & 96.4 & 15.3 & 78 &92.1&\textit{5.9}&83\\
                      & RJA-MMR                & \textbf{96.7} & \textbf{10.3} & \textbf{79} & \textbf{97.3} & \textbf{7.1} & \textbf{90} & \textbf{98.7} & \textbf{11.2} & \textbf{86}&  \textbf{100.0}& \textbf{4.3} & \textbf{92} \\
\hline
\multirow{10}{*}{Text-CNN} 
                      & BAE                    & 39.1 & 10.3 & 72.6 & 85.3 & 9.8   & 73  & 80.1 & 10.4 & 70 &71&7.9&80\\
                      & AGA                    & 33.3 & 11.3 & 75.4 & 81.1 & 7.7   & 85  & 77.4 & 10.8 & 75 &80&8.8&83\\
                      & FAGA                   & 56.1 & 11.5 & 80.1 & 90.1 & 8.3   & 80  & 92.1 & 15.3 & 69 &83&7.6&87\\
                      & MHA                    & 70.0 & 16.4 & 71.1 & 95.1 & 14.1  & 56  & 85.5 & 17.2 & 74 &91&10.1&81\\
                      & BA                     & 70.2 & 15.4 & 81.1 & 97.1 & 9.3   & 83  & 83.4 & 13.4 & 70 &89&8.7&80\\
                      & PWWS                   & 85.3 & 16.5 & 81.1 & 98.2 & 11.3  & 79  & 98.1 & 13.4 & 81 &\underline{99}&10.1&81\\
                      & TF                     & 77.3 & 19.4 & 74.9 & 91.5 & 10.9  & 83  & 91.0 & 17.2 & 75 &\underline{99}&\textit{6.9}&\underline{90}\\
                      & PSO                    & 76.2 & 15.5 & 77.3 & 99.0 & 9.4   & 83  & 92.2 & 17.2 & 81 &\underline{99}&9.3&88\\
                      & RJA                    & 88.3 & 11.4 & 79.1 & 94.9 & 9.7   & 83  & 94.0 & 17.2 & 77 &95&\underline{6.6}&\underline{90}\\
                      & RJA-MMR                & \textbf{93.8} & \textbf{9.9} & \textbf{82.2} & \textbf{99.3} & \textbf{7.4} & \textbf{89} & \textbf{99.3} & \textbf{10.3} & \textbf{82} &  \textbf{100.0}& \textbf{3.3} & \textbf{91}\\
\bottomrule
\end{tabular}
\label{tab: sar_mod_sim}
\end{table}

\begin{table}[ht]
\caption{Results on PPL and GErr metrics among the baselines and proposed methods on different datasets. The best performance is in bold.}
\small
\centering
\begin{tabular}{llcccccccc}
\toprule
\multirow{2}{*}{Task} & \multirow{2}{*}{Method} & \multicolumn{2}{c}{AG News} & \multicolumn{2}{c}{Emotion} & \multicolumn{2}{c}{SST2} & \multicolumn{2}{c}{IMDB
} \\ 
\cmidrule(lr){3-4} \cmidrule(lr){5-6} \cmidrule(lr){7-8} \cmidrule(lr){9-10}
                      &                         & PPL$\downarrow$ & GErr$\downarrow$ & PPL$\downarrow$ & GErr$\downarrow$ & PPL$\downarrow$ & GErr$\downarrow$ & PPL$\downarrow$ & GErr$\downarrow$ \\
\hline
\multirow{10}{*}{BERT-C} 
                      & BAE                    & 142   & 0.19 & 233 & 0.10 & 173 & 0.15 &181&0.21\\
                      & AGA                    & 132   & 0.21 & 291 & 0.14 & 192 & 0.17 &211&0.23\\
                      & FAGA                   & 165   & 0.19 & 259 & 0.13 & 182 & 0.24 &155&0.23\\
                      & MHA                    & 223   & 0.28 & 311 & 0.18 & 210 & 0.31 &160&\textit{0.20}\\
                      & BA                     & 281   & 0.19 & 301 & 0.17 & 200 & 0.25 &100&\textit{0.20}\\
                      & PWWS                   & 318   & 0.22 & 333 & 0.16 & 214 & 0.21 &\textit{89}&0.22\\
                      & TF                     & 312   & 0.28 & 341 & 0.19 & 191 & 0.17 &111&0.22\\
                      & PSO                    & 292   & 0.31 & 362 & 0.20 & 197 & 0.27 &91&0.22\\
                      & RJA                    & 155   & 0.21 & 281 & 0.12 & 201 & 0.18 &\underline{77}&\textbf{0.19} \\
                      & RJA-MMR                & \textbf{141} & \textbf{0.18} & \textbf{221} & \textbf{0.10} & \textbf{169} & \textbf{0.13} &\textbf{61}& \textbf{0.19}\\
\hline
\multirow{10}{*}{Text-CNN} 
                      & BAE                    & 132   & 0.19 & 201 & 0.10 & 143 & 0.13 &131&0.23 \\
                      & AGA                    & 132   & 0.13 & 213 & 0.10 & 182 & 0.13 &158&0.24\\
                      & FAGA                   & 145   & 0.15 & 241 & 0.13 & 163 & 0.17 &198&0.26\\
                      & MHA                    & 211   & 0.29 & 248 & 0.16 & 164 & 0.22 &156&\underline{0.22}\\
                      & BA                     & 241   & 0.15 & 221 & 0.13 & 210 & 0.21 &123&0.23\\
                      & PWWS                   & 277   & 0.20 & 299 & 0.19 & 224 & 0.23 &\underline{79}&\underline{0.22}\\
                      & TF                     & 199   & 0.21 & 314 & 0.21 & 194 & 0.21 &166&\underline{0.22}\\
                      & PSO                    & 142   & 0.14 & 301 & 0.18 & 145 & 0.14 &164&0.28\\
                      & RJA                    & 154   & 0.17 & 294 & 0.15 & 164 & 0.18 &\textit{89}&0.23\\
                      & RJA-MMR                & \textbf{127} & \textbf{0.12} & \textbf{190} & \textbf{0.08} & \textbf{142} & \textbf{0.11} &\textbf{65}&\textbf{0.19 }\\
\bottomrule
\end{tabular}
\label{tab: ppl_gerr}
\end{table}

\end{landscape}

\subsection{Experimental Results and Analysis}\label{main results}
The main experimental results of the attacking performance (SAR), the imperceptibility performance (Sim, Mod) and the fluency of adversarial examples (PPL, GErr) are listed in Table \ref{tab: sar_mod_sim} and \ref{tab: ppl_gerr}. Moreover, we demonstrate adversarial examples crafted by various methods shown in Table \ref{tab: example}. We manifest the three contributions mentioned in the Introduction section by answering three research questions:

\subsubsection*{Does our method make more thrilling attacks compared with baselines?}
We compare the attacking performance of the proposed method RJA-MMR and baselines in Table \ref{tab: sar_mod_sim}. This table demonstrates that RJA-MMR consistently outperforms other competing methods across different data domains, regardless of the structure of classifiers. Further, even RJA, by itself, without using MMR, can craft more menacing adversarial examples than most baselines. We attribute such an outstanding attacking performance to the two prevailing aspects of RJA. Firstly, RJA optimizes the performance by stochastically searching the domain. Most of the baselines perform a deterministic searching algorithm which could get stuck in the local optima. Differently, such a stochastic mechanism helps skip the local optima and further maximize the attacking performance.

Secondly, some of the baselines strictly attack the victim words in the order of word saliency rank (WSR), where the domain of the hierarchical search is limited to combinations of the neighboring victim words from the WSR, which would miss the potential optimal victim words combination. Unlike these methods, the RJA would enlarge the searching domain by testing more combinations of substitutions that do not follow the WSR order.  Thus, the proposed method RJA achieves the best-attacking performance, with the highest successful attack rate (SAR).

\begin{table}[t]
\centering
\caption{Adversarial examples of the Emotion dataset for victim classifier BERT-C. Blue texts are original words, while red ones are substitutions. Besides the examples, the attack performance is measured by attacking success and confidence in making correct predictions. The lower confidence indicates better performance and the successful attacks and lowest confidence are bold.}
\centering
\begin{tabular}{p{1.5cm}p{6cm}p{1.5cm}p{1.2cm}}
\toprule
Methods    & Adversarial example & Success & Confidence\\ \midrule
BAE        &       made a \del{wonderful} \nt{nasty} new friend         &\textbf{Successful} & 4.3\%     \\\midrule
AGA       &            made a \del{wonderful} \nt{beautiful} new friend   & Failed &   94\%  \\\midrule
FAGA        &            \del{made} \nt{introduced} a \del{wonderful} \nt{beautiful} new friend    &Failed&  95\%   \\\midrule
MHA        &             made a \del{wonderful} \del{new} \nt{newly} friend     &Failed&  70\% \\\midrule
BA     &           made a \del{wonderful} \nt{good} \del{new} \nt{brand} friend      &Failed&95\%    \\\midrule
PWWS       &           \del{made} \nt{seduce} a wonderful \del{new} \nt{raw} \del{friend} \nt{admirer} &Failed&   99\%   \\\midrule
TF &          made a \del{wonderful} \nt{strange} new friend       &\textbf{Successful}& 5.0\%   \\\midrule
PSO        &              \del{made} \nt{doomed} a wonderful new friend &\textbf{Successful}& 0.92\%      \\\midrule
RJA-MMR      &            made a \del{wonderful} \nt{lovely} new friend      &\textbf{Successful}& \textbf{0.80\%} \\\bottomrule
\end{tabular}
\label{tab: example}
\end{table}

\par
\subsubsection*{Is RJA-MMR superior to the baselines in terms of imperceptibility?}
We evaluate the imperceptibility of different attack strategies in terms of semantic similarities (USE) and modification rate (Mod) between the original input text and its derived adversarial examples, shown in Table \ref{tab: sar_mod_sim}. It can be seen that the proposed RJA-MMR attains the best performance among the baselines. The outstanding performance of the proposed method is attributed to the mechanisms of RJA and MMR. For semantic preservation, we statistically design the target distribution (Eq. \ref{eqt: target rja}) with a strong regularization of the semantic similarity in each iteration. Moreover, the HowNet is a knowledge-graph-based thesaurus that provides part-of-speech (POS) aware substitutions. Compared with the candidates supplied by baselines, the synonyms from HowNet can be more semantically similar to the original words. As for the modification rate, the proposed MMR is mainly designed for restoring the attacked words from successful adversarial examples so that the proposed RJA-MMR perturbs fewer words without sacrificing the attacking performance. Thus we can conclude that the proposed RJA-MMR provides the best performance for imperceptibility among baselines.

\subsubsection*{Is the quality of adversarial examples generated by the proposed methods better than that crafted by the baselines?} 
We insist the qualified adversarial examples should be parsing-fluent and grammarly correct. From the table \ref{tab: ppl_gerr}, we can find the RJA-MMR provides the lowest perplexity (PPL), which means the examples generated by RJA-MMR are more likely to appear in the corpus of evaluation. As our corpus is long enough and the evaluation model is broadly used, it indicates these examples are more likely to appear in natural language space, thus eventually leading to better fluency. For the grammar errors, the proposed method RJA-MMR is substantially better than the other baselines, which indicates a better quality of the adversarial examples. We attribute such performance to our method of finding word substitution, constructing the candidates set by intersecting the candidates from HowNet and MLM.

\subsection{Ablation Study}\label{ablation}
To rigorously validate the efficacy of the proposed RJA-MMR method, this section conducts a detailed ablation study, dissecting each component to assess its individual impact and overall contribution to the method's performance.
\subsubsection{Effectiveness of RJA}
We compare the attacking performance of our Reversible Jump Attack methods (RJA, RJA-MMR) and baselines in Table \ref{tab: sar_mod_sim}, reflected by SAR. The RJA helps attackers achieve the best attacking performance, with the largest metric SAR across the different downstream tasks. Apart from RJA-MMR, its ablation RJA also surpasses the strong baselines in most cases. Therefore, RJA is effective in terms of attacking performance. 

\subsubsection{Effectiveness of MMR}
MMR is a stochastic mechanism to reduce the modifications of adversarial examples with attacking performance preserved. Besides RJA-MMR, we also apply MMR to different attacking algorithms, including PSO, TF, PWWS, BA and MHA, aiming to demonstrate the advantages of MMR in general. 
\par
\begin{figure}[t]
    \centering
    \includegraphics[width=0.7\textwidth]{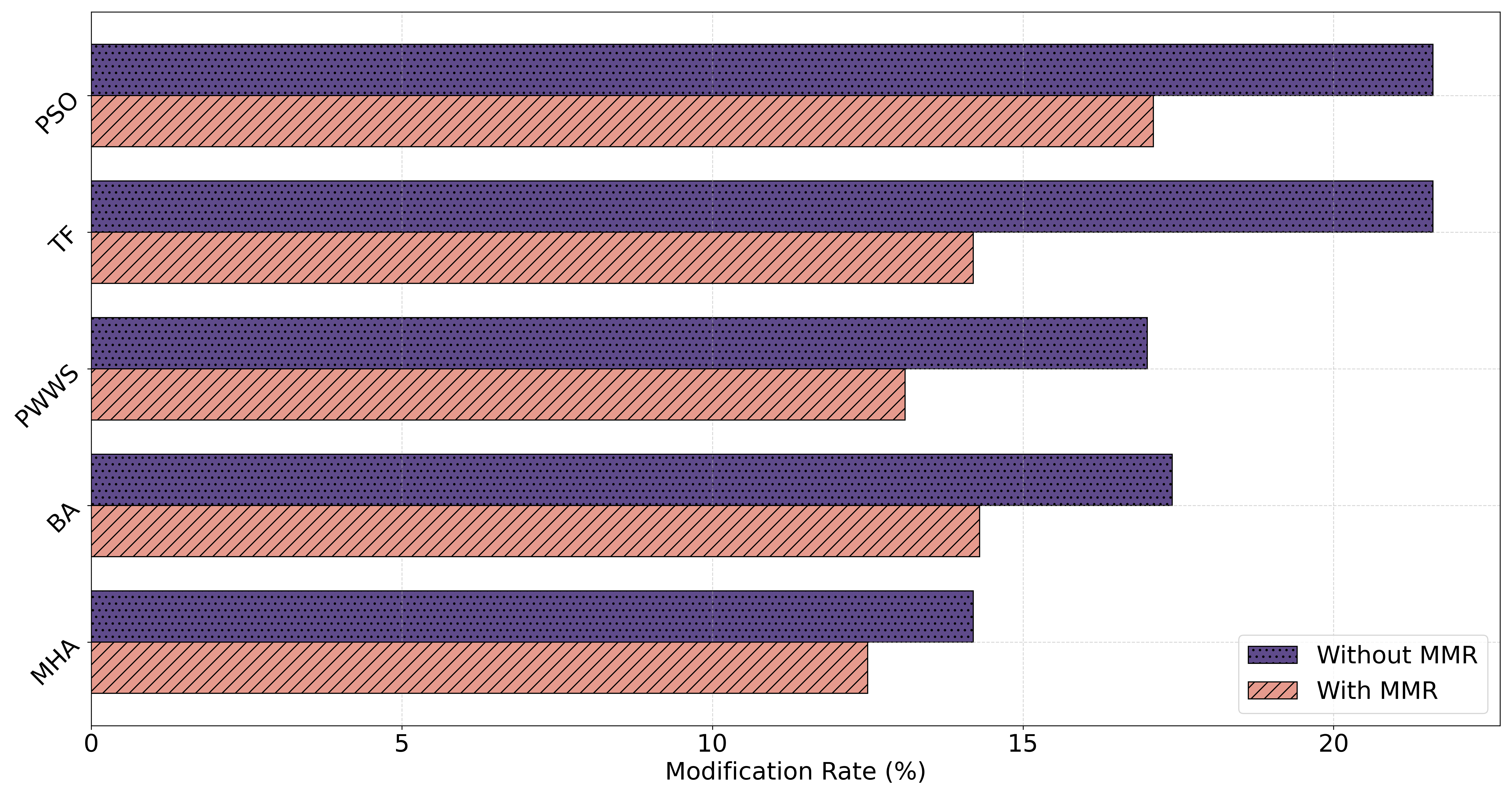}
    \caption{Comparisons on modification rates among attacking strategies (PSO, TF, PWWS, BA, MHA) with MMR and without MMR to attack the BERT-C on AG News dataset.}
    \label{fig: ablation mmr}
\end{figure}

From Table \ref{tab: sar_mod_sim}, we can find RJA-MMR has superior performance to RJA with lower modification rates. Moreover, the other baseline analysis results are shown in Fig \ref{fig: ablation mmr}. It shows that the attacking algorithms with MMR consistently have a lower modification rate than those without MMR. This means that attacking strategies can generally benefit from MMR by making fewer modifications.

\begin{table}[t]
    \centering
    \caption{Performance metrics for RJA-MMR against the TextCNN model on the AG News dataset using varied word candidate selection methods. The best performances for each metric are highlighted in bold.}
    \begin{tabular}{cccccc}
    \toprule
         Methods & SAR &  USE & Mod & PPL &GErr \\ \midrule
          HowNet & 85.1  & 72 & 13.1& 159 & 0.15\\   
          BERT-base & 83.1 & 70 & 12.1 & 144 & 0.15\\
          RoBERTa-large & 90.1 & 73 & 11.3 & 156 & \textbf{0.12}\\
          HowNet+BERT-base & 90.8 & 80 & 10.9 & 145 & 0.14\\ 
          \midrule
          HowNet+RoBERTa-large &\textbf{93.8} & \textbf{82} & \textbf{9.9} & \textbf{127} & \textbf{0.12}\\ 
          \bottomrule
    \end{tabular}
    \label{tab: ablation words}
\end{table}

\begin{table}[t]
    \centering
    \caption{Assessment of attack algorithms' efficiency on the Emotion dataset, utilizing empirical complexity (EC) in seconds per example for practical evaluation and total variance (TV) distance for theoretical convergence speed analysis. Lower EC values denote higher efficiency. The top three methods are highlighted in bold, italic, and underlined.}
    \begin{tabular}{p{1.2cm}p{0.5cm}p{0.5cm}p{0.5cm}p{0.5cm}p{0.5cm}p{0.75cm}p{0.5cm}p{0.55cm}p{0.5cm}p{0.85cm}}
    \toprule
        Methods & Metric&BAE & FAGA & MHA & BA & PWWS & TF & PSO & RJA & RJA-MMR \\ \midrule
    \multirow{2}{*}{BERT-C} & EC & \textit{21.7} &  162.4 & 414.0 & 707.9 & \textbf{0.7} & \underline{40.5} & 73.8 & 66.9 & 56.2  \\ \cmidrule{2-11}
    ~& TV& --&  \textit{1.22} & 1.14 & -- & -- & -- & 1.3 & \underline{0.99} & \textbf{0.89}  \\ \midrule
        \multirow{2}{*}{TextCNN} &EC & \underline{17.4} & 84.5 & 191.3 & 488.1 & \textbf{0.4} & \textit{28.1}& 55.1 & 51.9 & 54.1 \\ \cmidrule{2-11}
         ~& TV& --&  1.31 & 1.40 & -- & -- & -- & \textit{1.29} & \underline{1.11} &\textbf{ 1.01 } \\ 
        \bottomrule
    \end{tabular}
\label{tab: efficiency}
\end{table}
\subsubsection{Performance versus the Number of Iterations}
\begin{figure}[t!]
    \centering
    \includegraphics[width=0.95\textwidth]{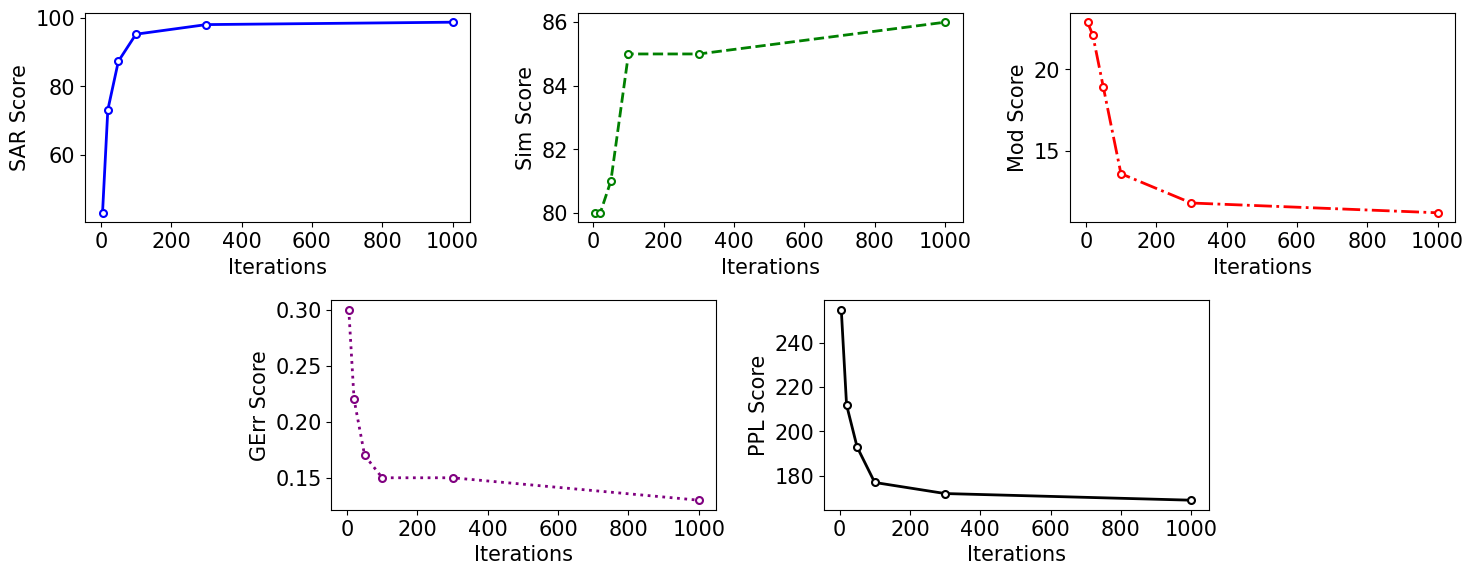}
    \caption{The progression of SAR, SIM, Mod, GErr, and PPL metrics for SST2 BERT over increased iterations (T). Performance trends and convergence points are visually represented.}
    \label{fig: T}
\end{figure}

The performance of the proposed methods is influenced by the number of iterations, denoted as \(T\). To delve deeper into this relationship, we conducted an extensive ablation study examining the correlation between performance and \(T\). Insights drawn from Figure \ref{fig: T} reveal a positive trend where performance amplifies in tandem with the number of iterations. Notably, performance begins to plateau, indicating convergence, at \(T=100\).

\subsubsection{Effectiveness of the Word Candidates}

In our ablation study, detailed in Table \ref{tab: ablation words}, we explored the effectiveness of various word candidate selection methods on the performance of RJA-MMR against the TextCNN model, utilizing the AG News dataset. Our evaluation included three strategies: using HowNet, MLMs with BERT-base \citep{Devlin2019BERTPO}, RoBERTa-large \citep{Liu2019RoBERTaAR}, and a synergistic approach combining HowNet and MLMs. Individually, HowNet and the MLM approaches showed notable performance, with RoBERTa-large slightly outperforming BERT-base. However, the combination of HowNet and MLMs produced superior results, surpassing the individual methods in all evaluated metrics, highlighting the significant advantage of integrating HowNet with MLMs to enhance the effectiveness of adversarial attacks.

Furthermore, our analysis of combination strategies for generating word candidates revealed that the more sophisticated MLM, RoBERTa-large, yielded a more effective attack performance than its less advanced counterpart, BERT-base. This finding suggests a positive correlation between advancements in MLM technology and enhancements in attack efficacy. We attribute this trend to the ability of more advanced MLMs to generate more relevant and suitable word candidates for use in attack methodologies, thereby increasing the precision and effectiveness of adversarial strategies.

\subsection{Platform and Efficiency Analysis}\label{efficiency}
In this section, we aim to evaluate the efficiency from both empirical and theoretical perspectives. To perform the empirical complexity (EV) evaluation, we carry out all experiments on RHEL 7.9 with the following specification: Intel(R) Xeon(R) Gold 6238R 2.2GHz 28 cores (26 cores enabled) 38.5MB L3 Cache (Max Turbo Freq. 4.0GHz, Min 3.0GHz) CPU, NVIDIA Quadro RTX 5000 (3072 Cores, 384 Tensor Cores, 16GB Memory) (GPU), and 88GB RAM. Table \ref{tab: efficiency} lists the time consumed for attacking BERT and TextCNN classifiers on the Emotion dataset. The metric of time efficiency is second per example, which means a lower metric indicates better efficiency. Results from Table \ref{tab: efficiency} show that our RJA and RJA-MMR run longer than some static counterparts (PWWS, BAE, TF) but are more efficient than the others, such as PSO, FAGA, MHA and BA. Nonetheless, the results of our methods running longer than some baseline methods indicate the genuine time needed to look for the more optimal adversarial examples.
\par
To theoretically gauge convergence speed, researchers employ the probabilistic concept of Mixing Time (MT), which denotes the duration for a Markov chain to approach its steady-state distribution closely \citep{kroesehandbook}. Given that MT is constrained by the total variation distance (TV) between the proposed and target distributions, TV is frequently used as a metric to quantify both the mixing time and speed of convergence\citep{metropolis1953equation, Green1995ReversibleJM}. Analysis of Table \ref{tab: efficiency} reveals that the proposed RJA-MMR method registers the lowest Total Variance (TV) distance, indicating superior theoretical performance in terms of convergence speed compared to other methods.

\subsection{Transferability}\label{transferability}
\begin{figure}[t]
    \centering
    \includegraphics[width=0.75\textwidth]{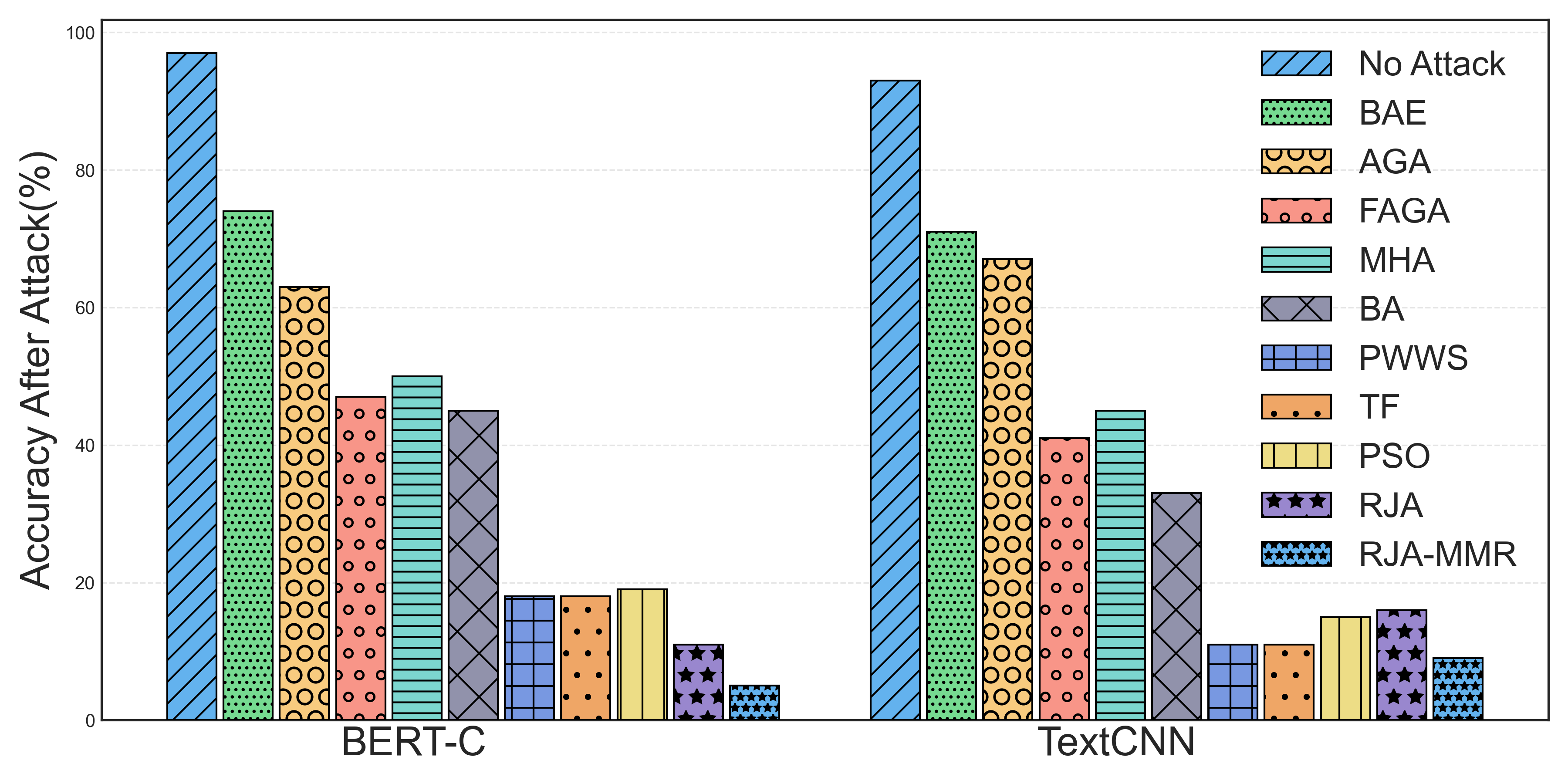}
    \caption{Performance of transfer attacks to victim models (BERT-C and TextCNN) on Emotion. A lower accuracy of the victim models indicates a higher transfer ability (i.e., the lower, the better).}
    \label{fig: transfer}
\end{figure}

The transferability of adversarial examples refers to its ability to degrade the performance of other models to a certain extent when the examples are generated on a specific classifier \citep{Goodfellow2015fgsm}. To evaluate the transferability, we investigate further by exchanging the adversarial examples generated on BERT-C and TextCNN and the results are shown in Fig \ref{fig: transfer}. 
\par
When the adversarial examples generated by our methods are transferred to attack BERT-C and TexCNN, we can find that the attacking performance of RJA-MMR still achieves more than 80\% successful rate, which is the best among baselines as illustrated in the Fig \ref{fig: transfer}.  Apart from RJA-MMR, its ablated components RJA also surpass the most baselines. This suggests that the transferring attacking performance of the proposed methods consistently outperforms the baselines.

\subsection{Targeted Attacks}\label{target}
A targeted attack is to attack the data sample with class $y$ in a way that the sample will be misclassified as a specified target class $y^{\prime}$ but not other classes by the victim classifier. RJA and MMR can be easily adapted to targeted attack by modifying $1-F_{y}(\mathbf{x})$ to  $F_{y^{\prime}}(\mathbf{x})$ in Eq. \ref{eqt: target rja}. The targeted attack experiments are conducted on the Emotion dataset. The results are shown in Table \ref{tab: target attack}, which demonstrates that the proposed RJA-MMR achieves better performance than PWWS, in terms of attacking performance (SAR), imperceptibility performance (Mod, Sim) and sentence fluency (GErr, PPL).

\begin{table}[t]
\caption{Targeted attack and imperceptibility-preserving performance on the Emotion dataset. The victim models are BERT-C and TextCNN classifiers, and the baseline is PWWS. The statistics for better performance are vertically highlighted in bold.}
\centering
\label{tab: target attack}
\begin{tabular}{cclllll}
\toprule
\multirow{2}{*}{Classifers} & \multicolumn{1}{c}{\multirow{2}{*}{Attack methods}} & \multicolumn{5}{c}{Metrics}                                                \\\cmidrule(lr){3-7}
                            & \multicolumn{1}{c}{}                                & SAR$\uparrow$           & Mod$\downarrow$           & PPL$\downarrow$         & GErr$\downarrow$          & Sim$\uparrow$         \\ \midrule
\multirow{2}{*}{BERT-C}       & PWWS                                                & 21.2          & 14.1          & 377          & 0.19          & 60          \\
                            & RJA-MMR                                             & \textbf{28.0}   & \textbf{9.2}  & \textbf{299} & \textbf{0.13} & \textbf{71} \\ \midrule
\multirow{2}{*}{TextCNN}    & PWWS                                                & 32.6          & 11.1          & 345          & 0.22          & 63          \\
                            & RJA-MMR                                             & \textbf{57.1} & \textbf{10.3} & \textbf{256} & \textbf{0.17} & \textbf{65} \\\bottomrule
\end{tabular}
\end{table}

\subsection{Attacking Models with Defense Mechanism}\label{defense}
Defending against textual adversarial attacks is paramount in ensuring the integrity and security of machine learning models used in natural language processing applications. Effective defense mechanisms encompass two multi-faceted approaches that include: 1) robust model training, utilizing adversarial training techniques to increase models’ resilience against malicious inputs. 2) malicious input detection, aiming to identify and mitigate adversarial examples without actively altering the machine learning model's structure or training process. 

To ensure a thorough evaluation of our proposed attack methods, we've integrated two distinct defense mechanisms into our assessment. For passive defense, we adopted the Frequency-Guided Word Substitutions (FGWS) \citep{passive} approach, which excels at identifying adversarial examples. Conversely, for active defense, we incorporated Random Masking Training (RanMASK)\citep{active}, a technique that bolsters model resilience via specialized training routines. We perform the adversarial attack to the BERT-C on the two datasets IMDB and SST2, and the results are presented in Table \ref{tab: defense}. The results show that our method outperforms the baselines.

\begin{table}[t]
\caption{A comparative analysis of attack performance (SAR) against BERT-C when subjected to two defense mechanisms, FGWS and RanMASK, across IMDB and SST2 datasets. Performance metrics are highlighted in bold to emphasize superior results.}
\centering
\label{tab: defense}
\begin{tabular}{ccllllll}
\toprule
Datasets &Defense &  BAE          &  FAGA           & MHA         & PWWS          & PSO & RJA-MMR         \\ \midrule
\multirow{2}{*}{IMDB}     & FGWS                                                & 37.7          & 18.0          & 34.9          & 66.1          & 80.0   &\textbf{88.1}       \\
                            & RanMASK                                             & 39.1   & 19.2  & 40.1 & 55.3 &81.0&\textbf{83.1} \\ \midrule
\multirow{2}{*}{SST2}    & FGWS                                                & 38.1          & 40.1          & 61.0          & 63.7          & 79.9 &\textbf{81.7}         \\
                            & RanMASK                                             & 41.1 & 16.4 & 39.6 & 71.3 & 77.1&\textbf{86.7} \\\bottomrule
\end{tabular}
\end{table}

\subsection{Adversarial Retraining}\label{retraining}
This section explores RJA-MMR's potential in improving downstream models' accuracy and robustness. Following \citep{li2021clare}, we use RJA-MMR to generate adversarial examples from AG's News training instances and include them as additional training data. We inject different proportions of adversarial examples into the training data for the settings of a BERT-based MLP classifier and a TextCNN classifier without any pre-trained embedding. We provide adversarial retraining analysis by answering the following two questions:
\begin{figure}[t]
    \centering
    \includegraphics[width=0.75\textwidth]{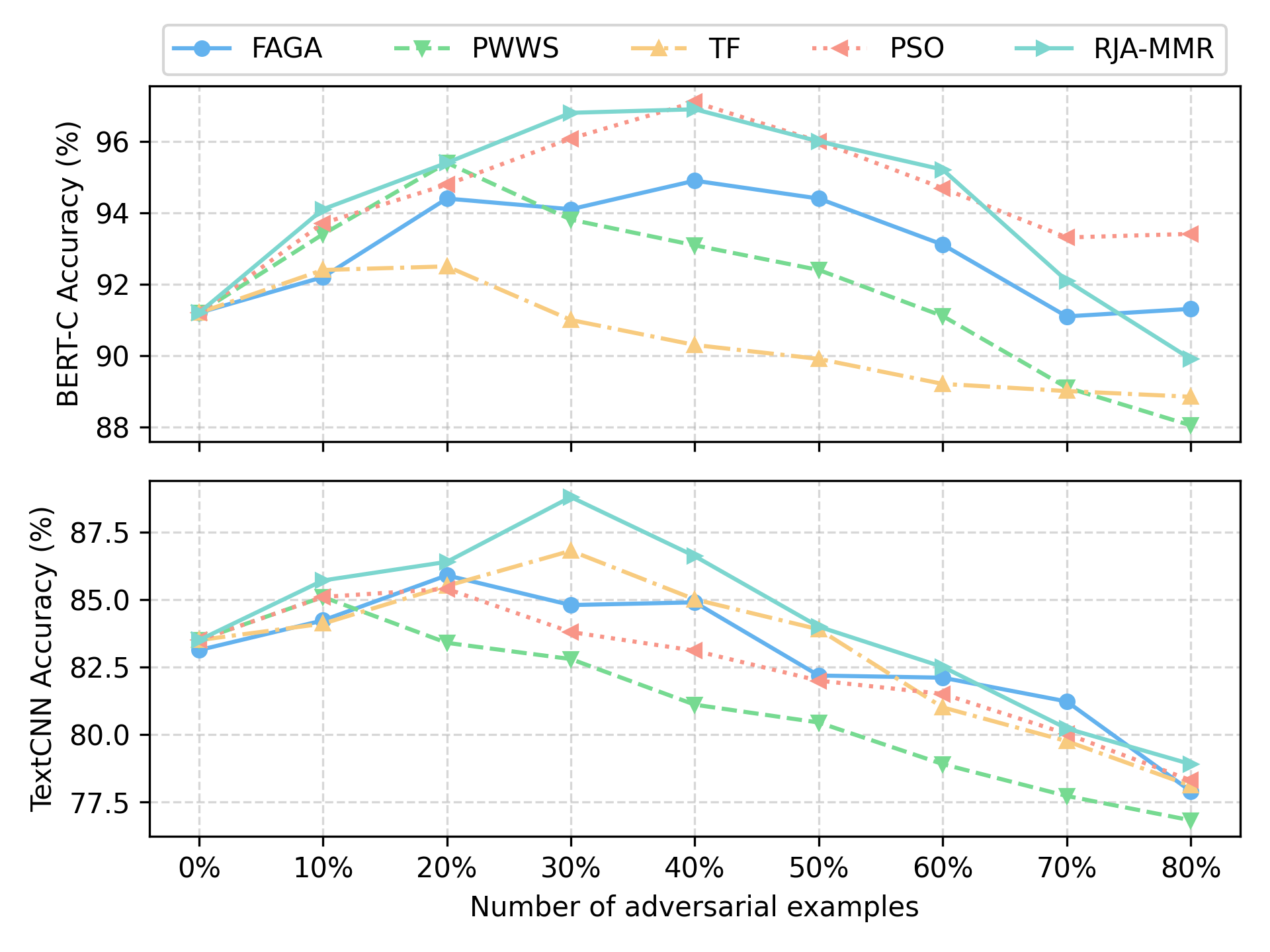}
    \caption{Results of adversarially trained BERT and TextCNN by inserting the different numbers of adversarial examples to the training set. The accuracy is based on the performance of the SST2 test set.}
    \label{fig:AdvTrain_accu}
\end{figure}

\subsubsection*{Can adversarial retraining help achieve better test accuracy?}
As shown in Fig. \ref{fig:AdvTrain_accu}, when the training data is accessible, adversarial training gradually increases the test accuracy while the proportions of adversarial data are smaller than roughly 30\%. Based on our results, we can see that a certain amount of adversarial data can help improve the models' accuracy, but too much such data will degrade the performance. This means that the right amount of adversarial data will need to be determined empirically, which matches the conclusions made from previous research \citep{jia2019faga,yang2021bigram}. 
\begin{figure}[t]
    \centering
    \includegraphics[width=0.65\textwidth]{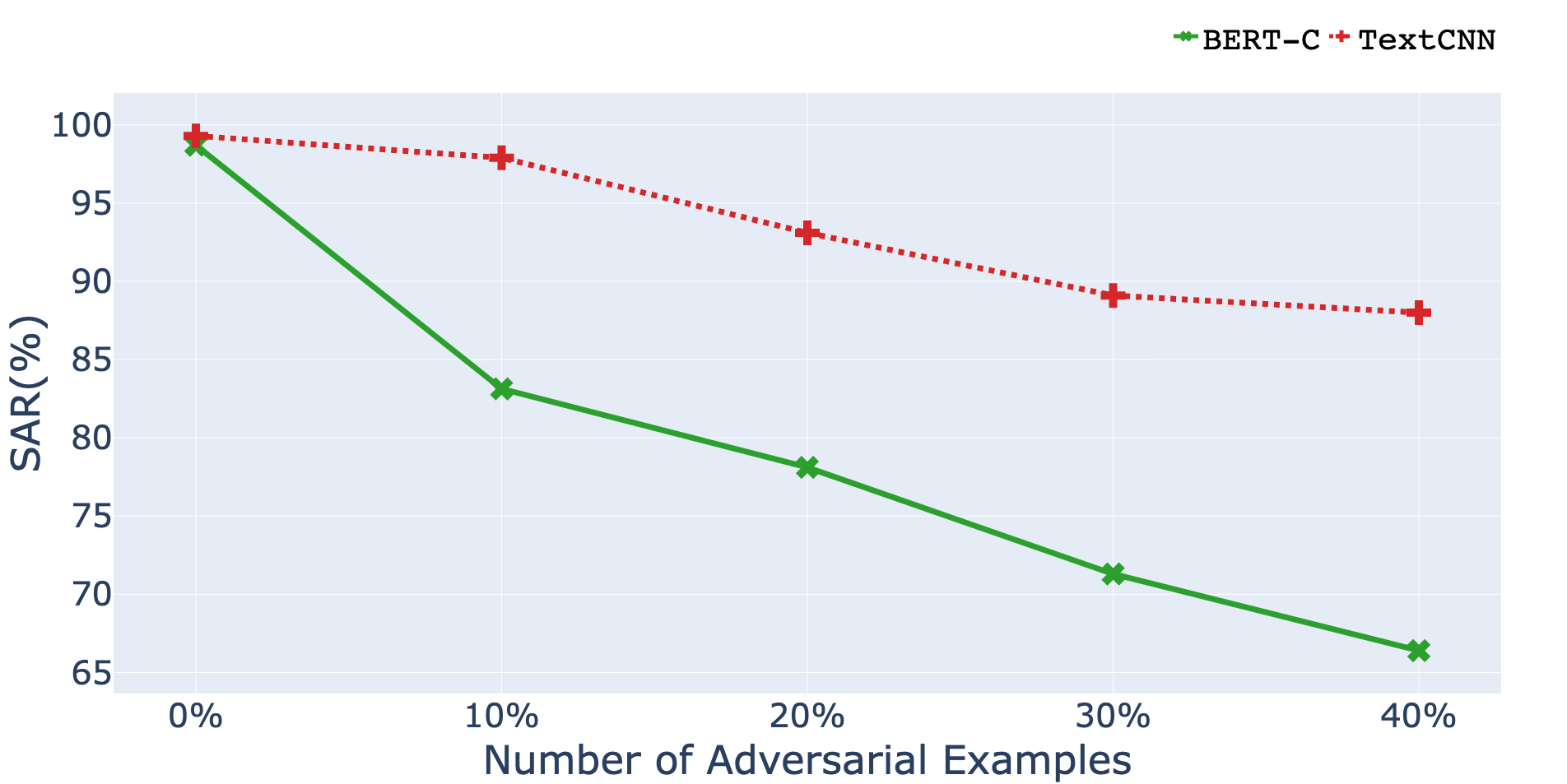}
    \caption{The success attack rate (SAR) of adversarially retrained models with different numbers of adversarial examples. A lower SAR indicates a victim classifier is more robust to adversarial attacks.}
    \label{fig:retraining_robust}
\end{figure}

\paragraph{Does adversarial retraining help the models defend against adversarial attacks?} To evaluate this, we use RJA-MMR to attack the classifiers trained with different proportions ($0\%,10\%,20\%,30\%,40\%$) of adversarial examples. A higher success rate (SAR) indicates a victim classifier is more vulnerable to adversarial attacks. As shown in Fig \ref{fig:retraining_robust}, adversarial training helps to decrease the attack success rate by more than 10\% for the BERT classifier (BERT-C) and 5\% for TextCNN. These results suggest that the proposed RJA-MMR can be used to improve downstream models' robustness by joining its generated adversarial examples to the training set.

\subsection{Parts of Speech Preference} \label{preference}
Regarding the superiority of the proposed method in attacking performance, we investigate its attacking preference, described by parts of speech (POS), for further linguistic analysis. In this subsection, we break down the attacked words in AG's News dataset by part-of-speech tags with Stanford PSO tagger \citep{POS}, and the collected statistics are shown in Table \ref{tab: POS}. By analyzing the results, we expect to find the more vulnerable POS by comparing the proposed methods and baselines.
\par
We apply PSO tagger to annotate them with POS tags, including \textit{noun}, \textit{verb}, \textit{adjective (Adj.)}, \textit{adverb (Adv.)} and \textit{others} (i.e., pronoun preposition, conjunction, etc.). Statistical results in Table \ref{tab: POS} demonstrate that all the attacking methods heavily focus on the \textit{noun}. Presumably, in the topic classification task, the prediction heavily depends on \textit{noun}. However, the proposed attacking strategies (RJA and RJA-MMR) tend to take a more significant proportion of \textit{others} than any other methods; thus we might conclude that \textit{Others} (pronoun, preposition and conjunction) might be the second adversarially vulnerable. Since these tags (pronouns, prepositions and conjunction) do not carry much semantics, we think these tags will not linguistically and semantically affect prediction but possibly impact the sequential dependencies, which could contaminate the contextual understanding of the classifiers and then subsequently cause wrong predictions.

\begin{table}[t!]
\caption{POS preference with respect to choices of victim words among attacking methods. The tags with the horizontally highest and second highest proportion are bold and italic, respectively.}
\centering
\begin{tabular}{p{1.5cm}|ccccc}
\toprule
Methods & Noun & Verb & Adj. & Adv. & Others \\
\midrule
BAE     & \textit{30\%}   & 14\%   & 13\%   & \textbf{41\%}   & 2\%      \\
AGA     & \textbf{44\%}   & \textit{21\%}   & 11\%   & 5\%    & 19\%     \\
FAGA    & \textbf{34\%}   & 11\%   & \textit{22\%}   & 14\%   & 19\%     \\
MHA     & \textbf{54\%}   & 9\%    & \textit{21\%}   & 4\%    & 12\%     \\
BA      & \textbf{68\%}   & 9\%    & 4\%    & 9\%    & \textit{10\%}     \\
PWWS    & \textbf{54\%}   & 9\%    & \textit{18\%}   & 3\%    & 16\%     \\
TF      & \textit{31\%}   & 10\%   & \textbf{39\%}   & 10\%   & 10\%    \\
PSO     & \textbf{48\%}   & 9\%    & 15\%   & \textbf{19\%}   & 9\%      \\
RJA     & \textit{28\%}   & 12\%   & 19\%   & 11\%   & \textbf{30\%}     \\
RJA-MMR & \textit{22\%}   & 17\%   & 13\%   & 17\%   & \textbf{31\%}  \\ \bottomrule
\end{tabular}
\label{tab: POS}
\end{table}

\begin{table}[t]
    \centering
    \caption{Robustness of BERT Models of Different Sizes on the Emotion Dataset. These models are trained with the same datasets and hyper-parameter but with different numbers of transformer layers (\textbf{L}) and hidden embedding sizes (\textbf{H}).}
    \label{tab: size pretrained}
    \begin{tabular}{lcccc}
        \toprule
        Versions & BERT Tiny & BERT Mini & BERT Small & BERT Medium \\
        \midrule
        Size & \textbf{L=2, H=128} & \textbf{L=2, H=128} & \textbf{L=4, H=512} & \textbf{L=4, H=512} \\
        \midrule
        SAR & 99.9 & 9.2 & 98.4 &  97.5\\
        Mod & 5.7 & 6.2 & 6.8 & 7.0 \\
        Sim & 93 & 92 & 91 & 91 \\
        \bottomrule
    \end{tabular}
\end{table}

\subsection{Robustness versus the Scale of Pre-trained Models}\label{scale}
Examining Tables \ref{tab: sar_mod_sim} and \ref{tab: ppl_gerr}, a question arises: Does increasing the scale of a model enhance its robustness? To explore this, we conducted a study applying our proposed attack methods to victim models of varying sizes on the Emotion dataset. 
\par
To provide a more nuanced analysis, we recognize that limiting our comparison to the two initial versions of BERT—base and large as introduced by \citep{Devlin2019BERTPO}—does not sufficiently support robust experimental outcomes. Hence, we have incorporated several widely recognized versions published subsequent to the original BERT paper. Specifically, we analyzed four versions of BERT as documented in \cite{tinybert}: BERT Tiny\footnote{\url{https://huggingface.co/prajjwal1/bert-tiny}}, BERT Mini\footnote{\url{https://huggingface.co/prajjwal1/bert-mini}}, BERT Small\footnote{\url{https://huggingface.co/prajjwal1/bert-small}}, and BERT Medium\footnote{\url{https://huggingface.co/prajjwal1/bert-medium}}. Notably, the most downloaded version among these has reached up to 6,559,486 monthly downloads on Huggingface alone. Our findings, detailed in Table \ref{tab: size pretrained}, demonstrate a positive correlation between model size and experimental robustness, confirming the value of incorporating a diverse range of model sizes into our analysis.

\section{Conclusion and Future Work}\label{sec13}

In recent years, the safety and fairness of NLP models have greatly been threatened by adversarial attacks. Many researchers have raised concerns about the robustness of the NLP classifiers because of their broad downstream tasks, such as fake news detection, sentiment analysis, and email spam detection. To improve classifiers' robustness, we have presented RJA-MMR which consists of two algorithms, Reversible Jump Attack (RJA) and Metropolish-Hasting Modification Reduction (MMR). RJA poses threatening attacks to NLP classifiers by applying the Reversible Jump algorithm to adaptively sample the number of perturbed words, victim words and their substitutions for individual textual input. While MMR is a customized algorithm to help improve the imperceptibility, especially to lower the modification rate, by utilizing the Metropolis-Hasting algorithm to restore the attacked words without affecting attacking performance. Experiments demonstrate that RJA-MMR delivers the best attack success, imperceptibility and sentence fluency among strong baselines.
\par
Although the adversarial examples can threaten the NLP models, these examples are not bugs but features \citep{ilyas2019adversarial}. To protect the models from the attacks, we conduct extensive experiments with a defense strategy, adversarial retraining, which is done by joining the adversarial examples in the training set and then retraining the models with the newly constructed training set. Unsurprisingly, in our experiments, the robustness of the classifiers has been greatly improved, while the accuracy of these models on clean data drops when an excessive amount of adversarial examples are injected. 

\par
Since the adversarial attack is one of the most effective methods to test the robustness of a model, the proposed attacks raise some concerns about deep neural networks (DNNs) and large pre-trained models. As DNNs and pre-trained language models achieved great success, most existing well-performed NLP classifiers are based on these techniques. Such popularity of these techniques could put textual classifiers at high risk because attackers can make effective attacks by utilizing DNNs and large pre-trained models. Thus a safer way of applying these techniques is a promising future research direction. At the same time, we also plan to pertinently study and design defense strategies to further improve the robustness of NLP classifiers under future adversarial attacks.
\backmatter





\section*{Declarations}
\begin{itemize}
\item Funding: Not Applicable.
\item Competing interests: Not Applicable.
\item Ethics approval: Not Applicable.
\item Consent to participate: The authors give their consent to participate.
\item Consent for publication: The authors give their consent to the publication of all information in this paper.
\item Availability of data and materials: All of the datasets are available on Huggingface (\url{https://huggingface.co/datasets}) and on our GitHub site (\url{https://github.com/MingzeLucasNi/RJA-MMR.git})
\item Code availability: All codes from our experiments are available at \url{https://github.com/MingzeLucasNi/RJA-MMR.git}
\item Authors' contributions: 
 Mingze Ni contributed to conceptualization, theoretical analysis, experiments and draft preparation; Zhensu Sun contributed to experiments, draft preparation, writing review; Wei Liu contributed to conceptualization, theoretical analysis, draft writing and editing.
\end{itemize}

\bibliography{sn-bibliography}


\end{document}